\documentclass[11pt,a4paper]{article}

\usepackage{jheppub_kim}
\usepackage{rotating}
\usepackage{graphicx,epsfig}
\usepackage{amsmath}
\usepackage {amssymb}
\usepackage{subfigure}

\usepackage{relsize}

\def\beq{\begin{equation}}
\def\eeq{\end{equation}}
\def\bea{\begin{eqnarray}}
\def\eea{\end{eqnarray}}

\begin{document}

\title{ Cosmological expansion and contraction from Pauli exclusion 
principle in  
$M0$-branes}

\author[a,b]{Alireza Sepehri}

\author[c]{Ahmed Farag Ali}

\author[d]{Kazuharu Bamba}

\author[e,f,g]{Salvatore Capozziello}

\author[h,i]{Richard Pincak}

\author[j]{Anirudh Pradhan}

\author[k]{Farook Rahaman}

\author[l,m,n]{Emmanuel N. Saridakis}

\affiliation[a]{Faculty of Physics, Shahid Bahonar University, P.O. Box 76175, Kerman,
Iran}

\affiliation[b]{Research Institute for Astronomy and Astrophysics of
Maragha (RIAAM), P.O. Box 55134-441, Maragha, Iran }

\affiliation[c]{Deptartment of Physics,
 Faculty of Science, Benha University, Benha 13518, Egypt}

\affiliation [d] {Division of Human Support System, Faculty of 
Symbiotic Systems Science, Fukushima University, Fukushima 960-1296, Japan}

\affiliation[e]{Dipartimento di Fisica, Universita' di Napoli Federico II, I-80126 - 
Napoli, Italy}

\affiliation[f]{INFN Sez. di Napoli, Compl. Univ. di Monte S. Angelo,
Edificio G, I-80126 - Napoli, Italy}

\affiliation[g]{Gran Sasso Science Institute (INFN), Viale F. Crispi, 7, I-67100
L'Aquila, Italy}

\affiliation[h]{Institute of Experimental Physics, Slovak Academy of Sciences,
Watsonova 47,043 53 Kosice, Slovak Republic}

\affiliation [i] {Bogoliubov Laboratory of Theoretical Physics, Joint
Institute for Nuclear Research, 141980 Dubna, Moscow region, Russia}

\affiliation[j]{Department of
 Mathematics, Institute of Applied Sciences and Humanities, G L A
 University, Mathura-281 406, Uttar Pradesh, India}

\affiliation [k] {Department of Mathematics, Jadavpur University, 
Kotkata - 700032, India}

\affiliation[l]{Instituto de F\'{\i}sica, Pontificia
Universidad de Cat\'olica de Valpara\'{\i}so, Casilla 4950,
Valpara\'{\i}so, Chile}

\affiliation[m]{CASPER, Physics Department, Baylor University, Waco, TX 76798-7310, USA}

\affiliation[n]{Physics Division,
National Technical University of Athens, 15780 Zografou Campus,
Athens, Greece}

\emailAdd{alireza.sepehri@uk.ac.ir}
\emailAdd{ahmed.ali@fsu.edu.eg}
 \emailAdd{bamba@sss.fukushima-u.ac.jp}
\emailAdd{capozziello@na.infn.it}
\emailAdd{pincak@saske.sk}
\emailAdd{pradhan@associates.iucaa.in}
\emailAdd{rahaman@associates.iucaa.in}
\emailAdd{Emmanuel$_-$Saridakis@baylor.edu}

\abstract{We show that the Pauli exclusion principle in a system of $M0$-branes 
can give rise to the expansion and contraction of the universe which is located on an 
$M3$-brane. We start with a system of  $M0$-branes with high symmetry, which join 
mutually and form pairs of $M1$-anti-$M1$-branes. The resulting symmetry breaking   
creates gauge fields that live on the $M1$-branes and play the role of graviton 
tensor modes, which induce an attractive force between the $M1$ and anti-$M1$ branes.
Consequently, the gauge fields that live on the $M1$-branes, and the scalar fields which 
are attached symmetrically to all parts of these branes, decay to fermions that attach 
anti-symmetrically to the upper and lower parts of the branes, and hence the Pauli 
exclusion principle emerges. By closing $M1$-branes mutually, the curvatures produced by 
parallel spins will be different from the curvatures produced by anti-parallel spins, and 
this leads to an inequality between the number of degrees of freedom on the boundary 
surface and the number of degrees of freedom in the bulk region. This behavior is 
inherited in the  $M3$-brane on which the universe is located, and hence this leads to 
the emergence of the universe  expansion and contraction.  In this sense, the Pauli exclusion principle rules the cosmic dynamics.}


\maketitle

\section{Introduction}

Recently, Padmanabhan has provided an explanation for the origin of the expansion of the 
universe, attributing it to the difference between the surface degrees of freedom and the 
bulk degrees of freedom \cite{Padmanabhan:2012ik}. Hence, several works were devoted on 
this novel approach and its various implications. In \cite{Yang:2012wn},
 the Friedmann equations of an ($n+1$)-dimensional Friedmann-Robertson-Walker 
(FRW) universe were obtained in the context of general relativity, Gauss-Bonnet 
gravity and Lovelock gravity (for reviews on modified gravity  see 
\cite{Nojiri:2006ri,Capozziello:2011et, Odintsov, Cai:2015emx}), while in  
\cite{Sheykhi:2013oac,Eune:2013ima,Chang-Young:2013gwa} these results were extended in 
the 
case of general spatial curvature. Additionally, in \cite{Ling:2013qoa} the Padmanabhan 
approach was generalized in the context of brane cosmology, scalar-tensor cosmology 
and $f(R)$ gravity, and in \cite{Ali:2013qza} it was incorporated in the framework of 
minimal length theories. 

The above approach can be also investigated in the context of BIonic systems, and it was 
suggested the the difference between surface degrees of freedom and bulk degrees of 
freedom could originate from the evolution of branes in extra dimensions 
\cite{Sepehri:2014jla}, where the BIon is defined as the configuration of two brane and 
anti-branes connected by a wormhole 
\cite{Sepehri:2015ara,Sepehri:2015eea,Sepehri:2016sjq,Sepehri:2016svv}. Moreover, the 
emergence and 
oscillation of cosmic space in a system of $M1$-branes has been investigated in detail in 
\cite{Sepehri:2016jfx}. Such a construction can be summarized as follows: Firstly, 
$M0$-branes join mutually to form a system of $M1$-anti-$M1$- branes. The 
scalar fields move only between $M0$-branes, since they are zero-dimensional objects, and 
hence these scalar fields could attach to them to play the role of scalar modes of the 
graviton. Consequently, with the birth of $M1$   and anti-$M1$ branes
\cite{Lozano:2000ara,Sepehri:2015ara,Sepehri:2016jfx}, gauge 
fields are produced, which have the role of graviton tensor modes, and lead to the 
formation of a wormhole between them. When the branes approach each others, the 
wormhole dissolves into them and leads to their expansion. By closing the branes 
mutually, the system square  energy becomes negative, and therefore the $M1$-branes needs 
to compactify, leading the the two-form gauge fields to convert to one-form gravity with 
opposite sign, and thus the sign of gravity reverses and anti-gravity arises. Finally, by 
joining $M1$-branes, $M3$-branes  \cite{Intriligator:2001ara, Cvetic:2002ara, 
Dhuria:2013ara,
 Kluson:2009ara} are created, which behave like the initial system. We 
mention that our universe is located on an  $M3$-brane, and by compactifying and opening 
it, the universe contracts and expands  \cite{Sepehri:2016jfx}. 
 
A question that arises naturally through the incorporation of Padmanabhan approach is 
how one could explain the process of fermion production and the origin of the Pauli 
exclusion principle, during the expansion of the universe. In the present work,  we will 
confront this question in the framework of a system of $M1$-anti-$M1$ branes. 

Initially, we consider that there exist only scalar fields that play the role of graviton 
scalar modes, and are attached to $M0$-branes. At this stage, the system has a high 
symmetry and no gauge fields or fermions live on the $M0$-branes. Then, $M0$-branes join  
mutually and form pairs of $M1$-anti-$M1$-branes. In this new stage, the symmetry of the 
system is broken, and gauge fields that live on the $M1$-branes arise and play the 
role of graviton tensor modes. Therefore, the attractive force between the $M1$ and 
anti-$M1$ branes leads to breaking of the symmetry of $M1$-branes, and thus the lower 
and upper parts of thee $M1$-branes are no longer the same. Consequently, the gauge 
fields that live on the $M1$-branes, and the scalar fields which are attached 
symmetrically to all parts of these branes, decay to fermions that attach 
anti-symmetrically to the upper and lower parts of the branes. Since fermions 
that live on the upper part of $M1$-branes have the opposite spin with respect to 
fermions that are placed on the lower part of the $M1$-branes, we deduce that when 
upper and lower spins are mutually linked $M1$-branes emerge and this implies the 
appearance of an attractive force between anti-parallel spins.

We mention that there is no link between parallel spins in $M1$-branes, and thus they 
cannot construct a stable system. This is exactly the reason for the emergence of the 
Pauli exclusion principle. By closing $M1$-branes mutually, the curvatures produced by 
parallel spins will be different from the curvatures produced by anti-parallel spins, and 
this leads to an inequality between the number of degrees of freedom on the boundary 
surface and the number of degrees of freedom in the bulk region, which according to 
Padmanabhan approach leads to the emergence of the expansion. Finally, according to the 
above discussion, since the  $M3$-branes  contract and open, the universe which is 
located on one of them experiences expansions and contractions. In other words, the 
universe evolution is a result of the difference between curvatures of parallel 
and anti-parallel spins and of the Pauli exclusion principle. 

The paper is organized as follows. In Section \ref{expcosmspace}, we consider the process 
of formation of fermions and an emergence of the Pauli exclusion principle in a system 
of  $M0$-branes. Additionally, in Section \ref{gravemerge}, we investigate the emergence of gravity due to   the relation between curvatures produced 
by identical and non-identical fermions. Differences between the number of degrees of 
freedom in the surface and in the bulk, showing that it can lead to the universe 
expansion, is discussed in Section \ref{freedom}. In Section \ref{Sectioncontraction}, we perform the analysis in the case 
of compactified branes, where the Pauli exclusion and the difference in number of degrees 
of freedom in the surface and in the bulk leads to universe contraction. Lastly, Section 
\ref{Conclusions} is devoted to the conclusions.
Details of calculations related  to the action terms and total actions are reported in the four final Appendices.

\section{The universe evolution from the emergence of the  Pauli exclusion 
principle}\label{expcosmspace}

 In this section we describe how one can attribute the universe evolution to  the Pauli 
exclusion principle. In particular, we consider four stages for the birth and 
the expansion of the universe. As we briefly described in the Introduction, at a first 
stage, there is a system with a high symmetry which includes only $M0$-branes and various 
scalars attached to them. At a second stage, by joining $M0$-branes and formating an 
$M1$-anti-$M1$-branes system, the symmetry of the system is broken and gauge fields 
emerge that play the role of gravitons. At a third stage, $M1$-branes interact with 
anti-$M1$-branes and the system symmetry is again broken. Under these conditions
the lower and upper parts of $M1$-branes are not the same, and thus gauge fields which 
live on $M1$-branes and scalars which are attached to them, decay to fermions with upper 
and lower spins that attach to upper and lower parts of the $M1$-branes respectively. 
This implies that each $M1$-brane is constructed from anti-parallel spins and hence there 
does not exist any stable object that can be built from parallel spins. Therefore, we 
deduce that there is going to appear an attractive force between anti-parallel spins and 
a repelling force between parallel spins, and the Pauli exclusion principle will emerge. 
Finally, at a fourth stage $M1$-branes come close to each other, and since the curvatures 
produced by parallel spins are different than the curvatures created by anti-parallel 
spins, there is going to appear a difference between the number of degrees of freedom on 
the boundary surface and the number of degrees of freedom in a bulk region, which 
according to Padmanabhan approach will trigger the universe expansion.


Let us start with the action of $M0$-branes that exist in the initial system. This reads
as 
\cite{Sepehri:2014jla,Sepehri:2016jfx,Sepehri:2015eea,Sepehri:2016sjq,
Sepehri:2016svv,Sepehri:2015ara}:
\begin{eqnarray}
S_{M0} = \int dt\, Tr\left\{ \Sigma_{M,N,L=0}^{10}
\langle[X^{M},X^{N},X^{L}],[X^{M},X^{N},X^{L}]\rangle\right\},
\label{s1}
\end{eqnarray}
where $X^{M}$(M=0,1,2,...10) are the scalars which are attached to $M0$-branes. They can 
be expressed as $X^{M}=X^{M}_{\alpha}T^{\alpha}$, with 
$T^{\alpha}$ the  generators  of the  Lie 3-algebra    
satisfying the 3-dimensional Nambu-Poisson brackets 
\cite{Bagger:2007jr,Gustavsson:2007vu,Ho:2008nn,Mukhi:2008ux}:
\begin{eqnarray}
 &&[T^{\alpha}, T^{\beta}, T^{\gamma}]= f^{\alpha \beta \gamma}_{\eta}T^{\eta} \nonumber 
\\&&\langle T^{\alpha}, T^{\beta} \rangle = h^{\alpha\beta} \nonumber \\&& 
[X^{M},X^{N},X^{L}]=[X^{M}_{\alpha}T^{\alpha},X^{N}_{\beta}T^{\beta},X^{L}_{\gamma}T^{
\gamma}]\nonumber \\&&\langle X^{M},X^{M}\rangle = X^{M}_{\alpha}X^{M}_{\beta}\langle 
T^{\alpha}, T^{\beta} \rangle,
\label{s2}
\end{eqnarray}
with $f^{\alpha \beta \gamma}_{\eta}$ the structure constants, and $h^{\alpha\beta}
$ the metric of system.

In order to obtain the total action of a $p$-dimensional system, one need to sum over 
actions of all $M0$-branes, resulting to:
\begin{eqnarray}
&&S_{system} = \int  d^{p+1}x \,\sum_{n=1}^{p}\beta_{n}\Big(
\delta^{a_{1},a_{2}...a_{n}}_{b_{1}b_{2}....b_{n}}L^{b_{1}}_{a_{1}}...L^{b_{n}}_{a_{n}}
\Big)^{1/2},
\end{eqnarray}
with
\begin{eqnarray}
&&\!\!\!\!\!\!\!
(L)^{a}_{b}=\delta_{a}^{b} \, Tr\Big\{  \Sigma_{a,b=0}^{p}\Sigma_{j=p+1}^{10}\Big[
\langle[X^{a},X^{i},X^{j}],[X_{a},X_{i},X_{j}]\rangle+
\langle[X^{a},X^{c},X^{d}],[X_{b},X_{c},X_{d}]\rangle
\nonumber\\
&& \ \ \ \ \ \ \ \ \ \ \ \ \ \ \ \ \ \ \ \ \ \ \ \ 
+
\langle[X^{a},X^{b},X^{
j}],[X_{b}
,X_{a},X_{j}]\rangle+
\langle[X^{k},X^{i},X^{j}],[X_{k},X_{i},X_{j}]\rangle
\Big]\Big\},
\label{s3}
\end{eqnarray}
where $\beta_{n}$ 
are constants, and where $a$, $b$ are indices of scalars on each brane, $i$, $j$ are 
indices of scalars in 
extra dimensions, and $p$ is the number of dimensions of the system.

By joining $M0$-branes and formating $M1$-branes, the symmetry is broken and gauge fields 
emerge. One can clearly see this using  the expressions 
\cite{Sepehri:2014jla,Sepehri:2016jfx,Sepehri:2015eea,Sepehri:2016sjq,
Sepehri:2016svv,Sepehri:2015ara,
Bagger:2007jr, Gustavsson:2007vu, Ho:2008nn, Mukhi:2008ux}:
 \begin{eqnarray}
 &&\!\!\!\!\!\!\!\!\!\!\!\!\!\!\!
 \langle[X^{a},X^{i},X^{j}],[X^{a},X^{i},X^{j}]\rangle= 
\frac{1}{2}\varepsilon^{abc}\varepsilon^{abd}(\partial_{c}X^{i}_{\alpha})(\partial_{d}X^{
i }_{\beta})
 \langle(T^{\alpha},T^{\beta})\rangle \sum_{j} (X^{j})^{2} 
\nonumber 
\\&& \ \ \ \ \ \ \ \ \ \ \ \ \ \ \  \ \ \ \ \ \  \ \ \ \ \ \ \ \ 
  =  \frac{1}{2}\langle \partial^{a}X^{i},\partial_{a}X^{i}\rangle \sum_{j} 
F(X^{j})\label{stt4} 
  \end{eqnarray}
  with $F(X)=\sum_{j} (X^{j})^{2}$, and where  $ i,j=p+1,..,10$, $\ a,b=0,1,...p$ and $
m,n=0,..,10$. Since
   \begin{eqnarray}
  &&\langle[X^{a},X^{b},X^{c}],[X^{a},X^{b},X^{c}]\rangle =-\lambda^{2}
 (F^{abc}_{\alpha\beta\gamma})(F^{abc}_{\alpha\beta\eta})\delta^{\kappa
 \sigma} \langle T^{\gamma},T^{\eta}\rangle= -\lambda^{2} \langle
 F^{abc},F^{abc}\rangle \label{st4}
  \end{eqnarray}
  with $F^{abc}_{\alpha\beta\eta}$ the strength fields in $M$-theory
  and   $\lambda^{2}
$ a constant, we can define
   \begin{eqnarray}
   && F_{abc}=\partial_{a} A_{bc}-\partial_{b}
 A_{ca}+\partial_{c}
 A_{ab}.
 \label{s4}
 \end{eqnarray}
 Hence, as we can see, we have effectively obtained the $2$-form gauge field  
$A_{ab}$, which is exchanged between branes, as a result of the  symmetry breaking that 
arose from the interaction between branes. Finally, inserting 
(\ref{stt4}), (\ref{st4}) and (\ref{s4}) into 
Eq. (\ref{s3}), we can re-write it as:
 \begin{eqnarray}
 &&\!\!\!\!\!\!\!\!\!\!\!\!\!\!   
 (L)^{a}_{b}=\delta_{a}^{b} Tr\Big\{  \Sigma_{a,b=0}^{p}\Sigma_{j=p+1}^{10}
 \Big[
    \frac{1}{2}\langle \partial_{a}X^{i},\partial_{a}X^{i}\rangle F (X^{j}) 
\nonumber\\
&& \ \ \ \ \ \ \ \ \ \ \  
+\frac{1}{
2}\langle \partial^{b}\partial^{a}X^{i},\partial_{b}\partial_{a}X^{i}\rangle-\lambda^{2} 
\langle
     F^{abc},F_{abc}\rangle +\langle[X^{k},X^{i},X^{j}],[X_{k},X_{i},X_{j}]\rangle
 \Big]\Big\},
 \label{s5}
 \end{eqnarray}
 which has the general form of an action in $M$-theory, namely
\cite{Sepehri:2014jla,Sepehri:2016jfx,Sepehri:2015eea,Sepehri:2016sjq,
Sepehri:2016svv,Sepehri:2015ara, 
Bagger:2007jr, Gustavsson:2007vu, Ho:2008nn, Mukhi:2008ux, Kluson:2000gr, Myers:1999ps, 
Tseytlin:1999dj,Constable:2001ag, Chu:2009iv, Sathiapalan:2008ye, Constable:1999ac}:
\begin{eqnarray}
  (L)^{a}_{b}=\delta_{a}^{b}  STr \left\{-det\left(P_{abc}[ E_{mnl}
  +E_{mij}(Q^{-1}-\delta)^{ijk}E_{kln}]+ \lambda
  F_{abc}\right)det(Q^{i}_{j,k})\right\}\label{ss5},
  \end{eqnarray}
where we make the identifications
 \begin{eqnarray}
  E_{mnl}^{\alpha,\beta,\gamma}
  &=& G_{mnl}^{\alpha,\beta,\gamma} + 
B_{mnl}^{\alpha,\beta,\gamma}, \nonumber\\
  Q^{i}_{j,k} &=& \delta^{i}_{j,k} + 
i\lambda[X^{j}_{\alpha}T^{\alpha},X^{k}_{\beta}T^{\beta},X^{k'}
_{\gamma}T^{\gamma}]
  E_{k'jl}^{\alpha,\beta,\gamma},
  \label{ss4}
 \end{eqnarray}
with
 $G_{mnl}=g_{mn}\delta^{n'}_{n,l}+\partial_{m}X^{i}\partial_{n'}X^{i}\sum_{j}
 (X^{j})^{2}\delta^{n'}_{n,l} + \frac{1}{2}\langle 
\partial^{b}\partial^{a}X^{i},\partial_{b}\partial_{a}X^{i}\rangle $, 
  $P_{abc}$   the pull-back of scalars, and where
$STr$ denotes the symmetric trace of products. Lastly, we can identify    
$\lambda=2\pi l_{s}^{2}$, with  $l_{s}$ is the string length in $M$-theory. 

Let us now consider the interaction between branes, which leads to breaking the symmetry 
again. Each scalar may decay to two fermions with upper and 
lower spins ($\psi^{U}$ and $\psi^{L}$ respectively), and each gauge field is 
exchanged between two fermions. We can then write \cite{Sepehri:2016sjq,Sepehri:2016svv}:
 \begin{eqnarray}
 && A_{ab}\rightarrow \psi^{U}_{a}\psi^{L}_{b}-\psi^{L}_{a}\psi^{U}_{b}\nonumber \\
  && X\rightarrow 
\psi^{U}_{a}A^{ab}\psi^{L}_{b}-\psi^{L}_{a}A^{ab}\psi^{U}_{b}+\Psi^{U}_{a}A^{ab}\Psi^{L}_{
b}-\Psi^{L}_{a}A^{ab}\Psi^{U}_{b}+\Psi^{U}_{a}A^{ab}\psi^{L}_{b}-\psi^{L}_{a}A^{ 
ab}\Psi^{U}
_{b}\nonumber \\
 &&\partial_{a} =\partial_{a}^{U}+\partial_{a}^{L} \nonumber \\
  &&\partial_{a}^{U}\psi^{U}_{a}=1,\;\partial_{a}^{L}\psi^{L}_{a}=1,
  \label{w4new}
 \end{eqnarray}
Using these definitions we can break the symmetry and create fermions. In particular, in 
this framework we may consider the gauge fields to be exchanged between fermions, such as 
the antisymmetric state will be produced. Previously, in 
\cite{Sepehri:2016sjq,Sepehri:2016svv}, 
it has been shown that by applying this definition the Pauli exclusion principle can be 
re-derived, and Dirac equations for fermions can be obtained. Specifically, 
using   expressions \ref{w4new}) one can make the splitting 
\cite{Sepehri:2016sjq,Sepehri:2016svv}:
 \begin{equation}
 \langle
 F^{abc},F_{abc}\rangle_{tot}\equiv\langle
 F^{abc},F_{abc}\rangle_{Free-Free}+\langle
 F^{abc},F_{abc}\rangle_{Free-Bound}+\langle
 F^{abc},F_{abc}\rangle_{Bound-Bound},
 \label{w5new}
 \end{equation}
where   ``Free-Free'' stands for the mutual interaction of two free fermions, 
``Free-Bound''  denotes the mutual interaction of free and bound fermions, and 
``Bound-Bound'' denotes the mutual interaction of two bound fermions. Thus, 
using the  definitions (\ref{w4new}), we can calculate the different terms of 
(\ref{ss5}) in terms of couplings of parallel and anti-parallel spins 
\cite{Sepehri:2016sjq,Sepehri:2016svv}. The 
corresponding expressions are shown in Appendix \ref{Apptermsnew}.  

In summary, from the above expressions it is clear that by breaking the symmetry of the 
brane, the usual equations of Dirac fields are obtained and the Pauli matrices are 
being produced, as well as the exact form of fermionic interaction. When $M0$-branes 
link to each other only gauge fields are produced, while if they join non-completely the 
symmetry of the system is broken. Hence, the upper and lower parts of the $M1$-brane that 
is constructed by joining $M0$-branes are different, and thus the fields that are 
attached to the upper part create fermions with upper spin, while those which are joined 
to lower part create fermions with lower spin \cite{Sepehri:2016sjq,Sepehri:2016svv}. 
Since the number of upper and lower spins in different brane points varies, the
coupling of anti-parallel spins is larger than the coupling of parallel spins, and 
consequently the curvature of anti-parallel spins is not eliminated by the curvature of 
parallel spins, and hence gravity in this brane is produced. Thus, each scalar 
decays to two anti-parallel spins, which are connected by a gauge field ($X\rightarrow 
\psi^{U}_{a}A^{ab}\psi^{L}_{b}-\psi^{L}_{a}A_{ab}\psi^{U}_{b}$). Some fermions are 
free, denoted by $\psi$, and some are bounded to each other and to the branes, denoted by 
$\Psi$. In summary, the reason for the appearance of an attraction force between fermions 
with anti-parallel spins is the production of a gravitational force between them (the
gauge fields that are produced play the role of the graviton tensor mode on the brane, 
namely $\Psi^{\dag a,U}\langle F_{abc},F^{i'bc}\rangle\psi_{i'}^{L}$).

\section{The emergence of gravity}
\label{gravemerge}

Recently, a new approach in the framework of $M$-theory appeared in the literature, 
which allows the construction of all $Mp$-branes from $M1$ and $M0$-branes 
\cite{Sepehri:2016jfx,Sepehri:2015eea,Sepehri:2016sjq,Sepehri:2016svv,Sepehri:2015ara}.
This mechanism suggests that only scalars live on $M0$-branes, however by linking 
$M0$-branes and formating $M1$-branes leads to the creation of gauge fields. These $M1$-
branes interact with anti-$M1$-branes and constitute a new system. For this system the
metric can be formed from the two $M1$-branes metrics, as
\begin{equation}
 \text{Metric of 
system}\equiv(\text{Metric M1})_{1}\otimes (\text{Metric M1})_{2}-(\text{Metric 
M1})_{2}\otimes (\text{Metric 
M1})_{1},
 \end{equation}
which implies that the system metric can be antisymmetric. Hence, since the graviton 
tensor mode has a direct relation with the metric, it can be anti-symmetric too.

Applying this approach, we  can calculate the relation between fermions and curvatures 
\cite{Sepehri:2016jfx,Sepehri:2015eea,Sepehri:2016sjq,Sepehri:2016svv,Sepehri:2015ara}:
\begin{eqnarray}
 &&A^{ab}=g^{ab}=h^{ab}=h_{1}^{ab'}\otimes h_{2}^{b'b}-h_{2}^{bb'}\otimes h_{1}^{ab'} 
 \nonumber\\&& 
 F_{abc}=\partial_{a} A_{bc}-\partial_{b}
 A_{ca}+\partial_{c} 
A_{ab}=2(\partial_{\mu}g_{\nu\lambda}+\partial_{\nu}g_{\mu\lambda}-\partial_{\lambda}g_{
\mu\nu})=
 2\Gamma_{\mu\nu\lambda}  
  \end{eqnarray}
  \begin{eqnarray}
 &&\!\!\!\!\!\!\!\!\!\!\!\!\!\!\!\!\!\!\!\!\!\!\!\!
 \langle
 F^{\rho}\smallskip_{\sigma\lambda},F^{\lambda}\smallskip_{\mu\nu}\rangle=
\langle[X^{\rho},X_{\sigma},X_{\lambda}],[X^{\lambda},X_{\mu},X_{\nu}]
\rangle
\nonumber\\
&&\ \ \ \ \ \  
=
[X_{\nu},[X^{\rho},X_{\sigma},X_{\mu}]]-[X_{\mu},[X^{\rho},X_{\sigma},X_{\nu}]]
\nonumber\\
&&\ \ \ \ \ \  \  \  \  \, +[X^{\rho}
,
X_{\lambda},X_{\nu}]
 [X^{\lambda},X_{\sigma},X_{\mu}]
-[X^{\rho},X_{\lambda},X_{\mu}][X^{\lambda},X_{\sigma},X_{\nu}]
\nonumber\\
 &&\ \ \ \ \ \   =
\partial_{\nu
}
\Gamma^{\rho}_{\sigma\mu}-\partial_{\mu}\Gamma^{\rho}_{\sigma\nu}+\Gamma^{\rho}_{
\lambda\nu}
 \Gamma^{\lambda}_{\sigma\mu}-\Gamma^{\rho}_{\lambda\mu}\Gamma^{\lambda}_{\sigma\nu}
 =R^{\rho}_{\sigma\mu\nu},
 \end{eqnarray}
where
$h^{ab}$ is the metric which is seen by the fermion, $X_{\sigma}$ is the scalar which is 
created by pairing two anti-parallel fermions, and $R^{\rho}_{\sigma\mu\nu}$ is the 
Riemann tensor. Additionally, we obtain
  \begin{equation}
  \langle
     F_{abc},F_{a'}^{bc}\rangle=R_{aa'}^{anti-parallel}-R_{aa'}^{parallel},
     \label{curvauxil}
      \end{equation}
      and
       \begin{eqnarray}
  &&
  R_{MN}=R_{
aa'}+R_{ia'}+R_{ij'}=R_{Free-Free}^{anti-parallel}+R_{Free-Bound}^{anti-parallel}+R_{
Bound-Bound}^{
anti-parallel}\nonumber\\
&&
\ \ \ \ \ \ \ \ \ \ \ 
-R_{Free-Free}^{parallel}-R_{Free-Bound}^{parallel}-R_{
Bound-Bound}^{
parallel}.
\label{w7new}
 \end{eqnarray}
 Finally, the term $\langle 
\partial^{b}\partial^{a}X^{i},\partial_{b}\partial_{a}X^{i}\rangle$ is given in Appendix 
\ref{Bigfermcurvtermnew}.

In summary, as we can see we obtain two types of gravity, one related to parallel spins 
and one related to anti-parallel spins, and  three types of curvatures, 
produced by the interaction of  free and bound fermions (the subscript ``Free-Free''
denotes the interaction of two free fermions that move between
branes, ``Free-Bound'' marks the interaction of one free fermion with one bound
fermion to $Mp$-branes, and ``Bound-Bound'' denotes the interaction of two
bound fermions to $Mp$-brane). Hence,  
$R_{Bound-Bound}^{anti-parallel}$ is the curvature produced by the interaction 
of two bound anti-parallel fermions, $R_{Bound-Bound}^{parallel}$ is the 
curvature produced by interaction of two bound parallel  fermions, 
$R_{Free-Free}^{anti-parallel}$ is the curvature created by the interaction of two free 
anti-parallel  fermions, $R_{Free-Free}^{parallel}$ is the curvature produced by 
interaction of two free parallel fermions, $R_{Free-Bound}^{anti-parallel}$ is the 
curvature created by the interaction of free and bound anti-parallel  fermions  and 
$R_{Free-Free}^{parallel}$ is the curvature produced by interaction of free and bound 
parallel  fermions. Lastly, we mention that the curvature between anti-symmetric fermions 
 will be positive and the curvature  between parallel spins is negative.

We can now use the above considerations in order to calculate the action of branes  
in (\ref{ss5}), in terms of curvature scalars and curvature tensors, following the 
approach of 
\cite{Sepehri:2016jfx,Sepehri:2015eea,Sepehri:2016sjq,Sepehri:2016svv,Sepehri:2015ara}. 
The details of the calculation are given in Appendix \ref{actioncalculation}, and the 
total   $Mp$-branes action writes as
 \begin{eqnarray}
 &&
 \!\!\!\!\!
 S_{Mp} = 
  - \int d^{p+1}x\sqrt{-g}
  \Biggl\{
  \Biggl\{
  \sum_{n=1}^{p}(1-m_{g}^{2})^{n}
 \Big[
 ( R_{Free-Free}^{parallel})^{2}+( R_{Free-Free}^{anti-parallel})^{2}+( 
R_{Free-Bound}^{parallel})
^{2}
\nonumber 
\\
&& \ \ \ \ \ \ \ \ \ \ \ \ \ \ \ \ \ \ \ \ \ \ \ \ 
+ ( R_{Free-Bound}^{anti-parallel})^{2}+( R_{Bound-Bound}^{parallel})^{2}+( 
R_{Bound-Bound}^{anti-
parallel})^{2}
\nonumber
\\
&& \ \ \ \ \ \ \ \ \ \ \ \ \ \ \ \ \ \ \ \ \ \ \ \ 
+
 (R_{Free-Free}^{parallel} 
R_{Free-Free}^{anti-parallel})\partial^{2}(R_{Free-Free}^{parallel}+ R_{
Free-Free}^{anti-parallel} )
\nonumber
\\
&& \ \ \ \ \ \ \ \ \ \ \ \ \ \ \ \ \ \ \ \ \ \ \ \ 
+(R_{Free-Bound}^{parallel} 
R_{Free-Bound}^{anti-parallel}
)\partial^{2}(R_{Free-Bound}^{parallel}+ R_{Free-Bound}^{anti-parallel} )
\nonumber 
\\
&& \ \ \ \ \ \ \ \ \ \ \ \ \ \ \ \ \ \ \ \ \ \ \ \ 
+(R_{Bound-
Bound}^{parallel} 
R_{Bound-Bound}^{anti-parallel})\partial^{2}(R_{Bound-Bound}^{parallel}+ 
 R_{Bound-Bound}^{anti-parallel} )
 \Big]^{n}
 \nonumber
 \\
 &&\ \ \ \ \ \
 + 
\sum_{n=1}^{p}m_{g}^{2n}\lambda^{2n}\delta_{\rho_{1}\sigma_{1}\cdots\rho_{n}\sigma_{n}}^{
\mu_
{1}\nu_{
1}\cdots
\mu_{n}\nu_{n}} 
\Big(
R^{anti-parallel, \rho_{1}\sigma_{1}}_{Free-Free,\mu_{1}\nu_{1}}+R^{anti-parallel,
\rho_ { 1
}\sigma_{
1}}_{Bound-Bound,\mu_{1}\nu_{1}}+ 
R^{anti-parallel, \rho_{1}\sigma_{1}}_{Free-Bound,\mu_{1}\nu_{1}}
\nonumber\\
&& \ \ \ \ \ \ \ \ \ \ \ \ \ \ \ \ \ \ \ \ \ \ \ \ \ \ \ \ \ \ \ \ \ \ \ \ \ \ \
-R^{
parallel, \rho_{
1}\sigma_{1}}_{Free-Free,\mu_{1}\nu_{1}}-
R^{parallel, \rho_{1}\sigma_{1}}_{Bound-Bound,\mu_{1}\nu_{1}}-R^{parallel, 
\rho_{1}\sigma_{1
}}_{Free-
Bound,\mu_{1}\nu_{1}}\Big)
\nonumber\\
&& \ \ \ \ \ \ \ \ \ \ \    \ \ \  \times\cdots
\cdots
\times
\Bigl(
R^{anti-parallel, \rho\sigma}_{Free-Free,\mu\nu}+R^{anti-parallel, \rho\sigma}_{
Bound-Bound,\mu\nu}+
 R^{anti-parallel, \rho_{n}\sigma_{n}}_{Free-Bound,\mu_{n}\nu_{n}}
 \nonumber\\
 &&  \ \ \ \ \ \ \ \ \ \ \  \ \ \ \ \ \ \ \ \ \  \ \ 
-R^{parallel, \rho_{n}\sigma_{n}}_{Free-Free,\mu_{n}\nu_{n}}-R^{parallel, 
\rho_{n}\sigma_{n}
 }_{Bound-
Bound,\mu_{n}\nu_{n}}-
R^{parallel, \rho_{n}\sigma_{n}}_{Free-Bound,\mu_{n}\nu_{n}}\Bigr)
\Biggr\}^{\frac{1}{2}}
\Biggr\}.
\label{s17}
 \end{eqnarray}
This  expression implies that the  total action of $Mp$-branes can be constructed from 
actions of $M1$-branes and depends on the curvatures of parallel and anti-parallel 
spins. If the curvatures of parallel spins are eliminated by the curvatures of 
anti-parallel spins, action (\ref{s17}) reaches its minimum value. By increasing and 
decreasing the difference between curvatures of parallel and anti-parallel spins, the 
action increases or decreases respectively. Furthermore, by increasing the mutual 
couplings of free and bound fermions, the action of the $Mp$-brane increases, too. 


In order to proceed, let us for simplicity choose 
 \begin{equation}
\Psi^{\dag a, 
L}\psi^{U}_{a}=\Psi^{\dag a, U}\psi^{L}_{a}=\Psi^{\dag a, L}\Psi^{U}_{a}=\Psi^{\dag a, 
U}\Psi^{L}_{a}=\psi^{\dag a, 
U}\psi^{L}_{a}=\psi^{\dag a, L}\psi^{U}_{a}=l_{1},
\label{choicepsi1new}
    \end{equation}
 and 
\begin{equation}
  \Psi^{\dag a, 
U}\psi^{U}_{a}=\Psi^{\dag a, 
L}\psi^{L}_{a}=\Psi^{\dag a, U}\Psi^{U}_{a}=\Psi^{\dag a, L}\Psi^{L}_{a}=\psi^{\dag a, 
U}\psi^{U}_{
a}=\psi^{\dag a, L}\psi^{L}_{a}=l_{2},
\label{choicepsi2new}
 \end{equation}
where $l_{1}$ is the separation distance between two $M1$-branes and $l_{2}$ is the 
length of $M1$-branes. $l_2$ has a direct relation with couplings of parallel 
spins, since  i) parallel spins are produced by   decay of a gauge field ($A_{ab}$)
which lives on an $M1$-brane and this field has a direct relation with the brane length 
\cite{Sepehri:2015ara,Sepehri:2016jfx}, and ii) since when a gauge field on each brane 
decays to two parallel spinors, the repulsive interaction between them causes them to 
turn away and thus move towards opposite brane sides, and therefore the length of the  
$M1$ brane increases. Additionally, $l_1$ has a direct relation with couplings of 
anti-parallel spins, since  i) anti-parallel spins are produced by   decay of a scalar 
field ($\psi$) which moves between $M1$-branes and has a direct relation with the 
separation distance between  branes \cite{Sepehri:2015ara,Sepehri:2016jfx}, and ii) since 
when a scalar field on each brane decays to two anti-parallel spinors, the attractive 
interaction between them causes the branes to approach each other, and thus the 
separation distance decreases.

When the branes are placed at a large 
distance 
from each other 
(i.e $l_{1}\rightarrow \infty$) the upper spin of the upper part of the $M1$-brane 
attracts the lower spin of the lower part of it, and therefore the length of  the 
$M1$-brane decreases (i.e. $l_{2}\rightarrow 0$). On the other hand, by bringing the  
$M1$-branes mutually closer (i.e $l_{1}\rightarrow 0$) the interaction between the upper 
and lower parts of two  $M1$-branes increases, comparing to the interaction of lower with 
upper spins, and therefore the length of $M1$-branes  increases (i.e $l_{2}\rightarrow 
\infty$).

Inserting the ansatzes (\ref{choicepsi1new}),(\ref{choicepsi2new}) into expressions 
(\ref{apprel1}), (\ref{apprel2}), (\ref{apprel3}), (\ref{apprel4}) and (\ref{w8}), we can 
calculate the various curvatures as
\begin{eqnarray}      
&&R_{Free-Free}^{parallel}=R_{Free-Bound}^{parallel}=R_{Bound-Bound}^{parallel}\approx 
l_{2}-
l'_{2}
\label{Rl1l2a} 
\\&&R_{Free-Free}^{anti-parallel}=R_{Free-Bound}^{anti-parallel}=R_{Bound-Bound}^{
anti-parallel}\approx l_{1}+l'_{1},
\label{Rl1l2b}
\end{eqnarray}
and by inserting these results into (\ref{s17}) we find 
the simple action:
\begin{eqnarray}
   && 
   \!\!\!\!\!\!\!\!\!\!\!\!\!\!\!\!\!
   S_{Mp-brane}\approx V\int  dt\,\Sigma_{n=1}^{p} 
\Big\{6 m_{g}^{2}\lambda^{2}\left[l_{1}-
l_{2}-l'_{1}+l'_{2}+(l'_{1})^{2}-(l'_{2})^{2}\right]
\nonumber
\\
&& \ \ \ \ \ \ \ \ \  \ \ \ \ \ \ \ \  \ \ \ \ \ \  
-3(1-m_{g}^{2})
\left[2l_{1}^
{2}+2l_{2}
^{2}+2(l'_{1})^{2}+2(l'_{2})^{2}+l_{1}^{2}l_{2}^{2}(l_{1}^{2}+l_{2}^{2})''\right] 
\Big\}^{\frac{n}{2}},
\label{w13}
       \end{eqnarray}
with $V$  the space-volume of the $Mp$-brane, and where  we have assumed that the 
couplings  depend only on time $t$, with $'$ denoting derivative with respect to 
$t$.

Variation of action (\ref{w13}) leads to the equations of motion:
\begin{eqnarray}      
&&
\!\!\!\!\!\!\!\!\!\!\!\!\!\!
\Sigma_{n=1}^{p}
\Biggl\{l'_{1}  \left\{[m_{g}^{2}(\lambda^{2}+1)-1](1+2l_{1}^{2}l_{2}^{
2} )\right\}      
\Big\{
6m_{g}^{2}\lambda^{2}[l_{1}^{2}-l_{2}^{2}+(l'_{1})^{2}-(l'_{2})^{2}]
\nonumber\\
&& \ \ \ \ \ \ \ \ \ \ \ \ \ \ \ \
-3(1-m_{g}^{2})[2l_{1}^{2}+2l_{2}^2+2(l'_{1})^{2}+2(l'_{2})^{2}+l_{1}^{2}l_{2}^{2}(l_{1
}^{
2}+l_{2}^{
2})'']
\Big\}^{\frac{n}{2}-1}\Biggr\}'
\nonumber\\
&&      
\!\!\!\!\!\!\!\!\!\!\!\!\!\!
=
\Sigma_{n=1}^{p}
\Biggl\{
l_{1}
\left\{
[m_{g}^{2}(\lambda^{2}+1)-1]
\left[1+3l_{2}^{2}(l_{1}
^{2}
+l_{2}^{2})''\right]\right\}      
\Big\{6m_{g}^{2}\lambda^{2}[l_{1}^{2}-l_{2}^{2}+(l'_{1})^{2}-(l'_{2})^{2}]
\nonumber\\
&& \ \ \ \ \ \ \ \ \ \ \ \ \ \ \ \       
-3(1-m_{g}^{2})[2l_{1}^{2}+2l_{2}^{2}+2(l'_{1})^{2}+2(l'_{2})^{2}+l_{1}^{2}l_{2}^{2}(l_{1
}
^{2}
+l_{2}^{2})'']\Big\}
^{\frac{n}{2}-1}
\Biggr\},
\label{s20}
\end{eqnarray}
and
\begin{eqnarray}      
&&
\!\!\!\!\!\!\!\!\!\!\!\!\!\!
\Sigma_{n=1}^{p}
\Biggl\{l'_{2}  \left\{[m_{g}^{2}(\lambda^{2}+1)-1](1-2l_{1}^{2}l_{2}^{
2} )\right\}      
\Big\{
6m_{g}^{2}\lambda^{2}[l_{1}^{2}-l_{2}^{2}+(l'_{1})^{2}-(l'_{2})^{2}]
\nonumber\\
&& \ \ \ \ \ \ \ \ \ \ \ \ \ \ \ \
-3(1-m_{g}^{2})[2l_{1}^{2}+2l_{2}^2+2(l'_{1})^{2}+2(l'_{2})^{2}+l_{1}^{2}l_{2}^{2}(l_{1
}^{
2}+l_{2}^{
2})'']
\Big\}^{\frac{n}{2}-1}\Biggr\}'
\nonumber\\
&&      
\!\!\!\!\!\!\!\!\!\!\!\!\!\!
=
\Sigma_{n=1}^{p}
\Biggl\{
l_{2}
\left\{
[m_{g}^{2}(\lambda^{2}+1)-1]
\left[1-3l_{1}^{2}(l_{1}
^{2}
+l_{2}^{2})''\right]\right\}      
\Big\{6m_{g}^{2}\lambda^{2}[l_{1}^{2}-l_{2}^{2}+(l'_{1})^{2}-(l'_{2})^{2}]
\nonumber\\
&& \ \ \ \ \ \ \ \ \ \ \ \ \ \ \ \       
-3(1-m_{g}^{2})[2l_{1}^{2}+2l_{2}^{2}+2(l'_{1})^{2}+2(l'_{2})^{2}+l_{1}^{2}l_{2}^{2}(l_{1
}
^{2}
+l_{2}^{2})'']\Big\}
^{\frac{n}{2}-1}
\Biggr\}.
\label{s21}
\end{eqnarray}
We can easily extract the approximate solution of these two equations as:  
\begin{eqnarray}
 &&l_{1}(t)\approx 
\Sigma_{n=1}^{p}
\left(1-\frac{t}{t_{s}}+m_{g}^{2}\right)^{\frac{n}{2}+2}e^{m_{g}^{2}\lambda^{2}
(1-\frac{t}{t_{s}}+m_{g}^{2})}
 \label{s22a}
 \\&& 
        l_{2}(t)
        \approx 
\Sigma_{n=1}^{p}
\left(1-\frac{t}{t_{s}}+m_{g}^{2}\right)^{-\frac{n}{2}-2}
e^{-m_{g}
^{2}\lambda^{2}(1-\frac{t}{t_{s}}+m_{g}^{2})}
       \left [1+ \left(1-\frac{t}{t_{s}}+m_{g}^{2}\right)^{2}\right].
 \label{s22}
\end{eqnarray}
In the above expressions $t=0$ corresponds to the Big bang time, and (for small 
$m_{g}$) $t_s$ is the colliding moment of the branes. Due to the presence of the 
exponential in solutions (\ref{s22a}),(\ref{s22}), we deduce that as time 
passes  $Mp$ and anti-$Mp$-branes come mutually closer, and the coupling between 
anti-parallel spins $l_{1}(t)$ decreases and tends to zero, while the coupling between 
parallel spins $l_2(t)$ grows and tends to infinity. Hence, due to Pauli exclusion 
principle, parallel spins repel and turn away mutually, the length of the $M1$-brane 
increases, and this corresponds to universe (which lies on an $M3$-brane 
\cite{Cvetic:2002ara,Kluson:2009ara} which has been constructed from $M1$-branes 
\cite{Lozano:2000ara,Sepehri:2016jfx}) expansion.

\section{The number of degrees of freedom in the boundary and in the  bulk}
\label{freedom}

Let us now calculate the number of degrees of freedom in the boundary and in the  bulk.
For simplicity, we first calculate the number of degrees of freedoms for the $M1$-brane 
in terms of couplings of parallel and anti-parallel spins, and then we obtain the total 
number of $Mp$-branes by summing over these numbers. 

It is clear that the sum of degrees of freedom in the bulk and on the surface 
of an $M1$-brane has a direct relation with the total energy of the system, namely 
\cite{Sepehri:2016jfx}:
\begin{eqnarray}
&&\!\!\!\!\!\!\!\!
N_{sur}^{M1}+N_{bulk}^{M1}\approx E_{system}=\nonumber 
\\&&  \   \
=     \int d^{2}x   
\sqrt{-g}\,
      \Bigg\{\!
     (1-m_{g}^{2})
                \Big[( R_{Free-Free}^{parallel})^{2}+( 
R_{Free-Free}^{anti-parallel})^{2}+( 
R_{Free-
Bound}^{parallel})^{2} 
\nonumber 
\\
&& \ \ \ \ \ \ \ \ \ \ \ \ \ \ \ \ \  \  \ \  \ \
+
( R_{Free-Bound}^{anti-parallel})^{2}+( 
R_{Bound-Bound}^{
parallel})^{2}+( R_{Bound-Bound}^{anti-parallel})^{2}
\nonumber
\\
&& \ \ \ \ \ \ \ \ \ \ \ \ \ \ \ \ \  \  \ \  \ \ + 
(R_{Free-Free}^{parallel} R_{
Free-Free}^{anti-parallel})\,\partial^{2}(R_{Free-Free}^{parallel}+ 
R_{Free-Free}^{anti-parallel} )
\nonumber \\
&& \ \ \ \ \ \ \ \ \ \ \ \ \ \ \ \ \  \  \ \  \ \
+(R_{Free-Bound}^{parallel} 
R_{Free-Bound}^{anti-parallel})\,\partial^{2}(R_{Free-Bound}^{
parallel}+ R_{Free-Bound}^{anti-parallel} )
\nonumber \\
&& \ \ \ \ \ \ \ \ \ \ \ \ \ \ \ \ \  \  \ \  \ \
+
(R_{Bound-Bound}^{parallel} 
R_{Bound-
Bound}^{anti-parallel})\,\partial^{2}(R_{Bound-Bound}^{parallel}+ 
R_{Bound-Bound}^{anti-parallel} ) \Big]
\nonumber\\
&& \ \ \ \ \ \ \ \ \ \ \ \ \ \ \ \ \  \ \
-  m_{g}^{2}\lambda^{2}\delta_{\rho_{1}\sigma_{1}}^{\mu_{1}\nu_{1}}              
\Big[
R^{anti-parallel, \rho_{1}\sigma_{1}}_{Free-Free,\mu_{1}\nu_{1}}+R^{anti-parallel,
\rho_{1}\sigma_{1}}_{Bound-Bound,\mu_{1}\nu_{1}}+R^{anti-parallel, \rho_{1}\sigma_{1}}_{
Free-Bound,\mu_{1}\nu_{1}}
\nonumber\\
&& \ \ \ \ \ \ \ \ \ \ \ \ \ \ \ \ \  \  \ \  \ \
-
R^{parallel, \rho_{1}\sigma_{1}}_{Free-Free,\mu_{
1}\nu_{1}} -R^{
parallel, \rho_{1}\sigma_{1}}_{Bound-Bound,\mu_{1}\nu_{1}}-R^{parallel, 
\rho_{1}\sigma_{1}}_
{Free-
Bound,\mu_{1}\nu_{1}}
\Big]  
\Bigg\}.
\label{T1}
\end{eqnarray}
Additionally, the difference between the number of degrees of freedom in the bulk and on 
the surface of an $M1$-brane leads to a change in the energy of the system, and to the 
emergence of a force between $M1$-branes 
\cite{Sepehri:2014jla,Sepehri:2016jfx,Sepehri:2016sjq,Sepehri:2016svv}. Hence, 
assuming for simplicity that all curvatures of parallel spins exhibit the same behavior, 
as well as all curvatures of anti-parallel spins, we can write:
\begin{eqnarray}      
&&\!\!\!\!\!\!\!\!\!\!\!
N_{sur}^{M1}-N_{bulk}^{M1}\approx
\frac{\partial E_{system}}{\partial t}=\frac{\partial E_{system}}{\partial 
l_{1}}\frac{\partial l_{1}}{\partial 
t}+\frac{\partial E_{system}}{\partial l_{2}}\frac{\partial l_{2}}{\partial t}
\nonumber 
\\&&
=
  \int d^{2}x   \sqrt{-g}\,
      \Bigg\{    
      (1-m_{g}^{2})
         \Big[( R_{Free-Free}^{parallel})^{2}-( 
R_{Free-Free}^{anti-parallel})^{2}+( 
R_{Free-
Bound}^{parallel})^{2}
\nonumber \\
&& \ \ \ \ \ \ \ \ \ \ \ \ \ \  \ \  \ \  \ \
-
( R_{Free-Bound}^{anti-parallel})^{2}+( 
R_{Bound-Bound}^{
parallel})^{2}-( R_{Bound-Bound}^{anti-parallel})^{2}
\nonumber \\
&& \ \ \ \ \ \ \ \ \ \ \ \ \ \  \ \  \ \  \ \
+ 
(R_{Free-Free}^{parallel} R_{
Free-Free}^{anti-parallel})\,\partial^{2}(R_{Free-Free}^{parallel}- 
R_{Free-Free}^{anti-parallel} )
\nonumber 
\\
&& \ \ \ \ \ \ \ \ \ \ \ \ \ \  \ \  \ \  \ \
+
(R_{Free-Bound}^{parallel} 
R_{Free-Bound}^{anti-parallel})\,\partial^{2}(R_{Free-Bound}^{
parallel}- R_{Free-Bound}^{anti-parallel} )
\nonumber
\\
&& \ \ \ \ \ \ \ \ \ \ \ \ \ \  \ \  \ \   \ \
+
(R_{Bound-Bound}^{parallel} 
R_{Bound-
Bound}^{anti-parallel})\,\partial^{2}(R_{Bound-Bound}^{parallel}- 
R_{Bound-Bound}^{anti-parallel} )
 \Big]
\nonumber\\
&& \ \ \ \ \ \ \ \ \ \ \ \ \ \   \ \ \ \
-
m_{g}^{2}\lambda^{2}\delta_{\rho_{1}\sigma_{1}}^{\mu_{1}\nu_{1}}        
   \Big[ R^{anti-parallel, \rho_{1}\sigma_{1}}_{Free-Free,\mu_{1}\nu_{1}}+R^{anti-
parallel, \rho_{1}\sigma_{1}}_{Bound-Bound,\mu_{1}\nu_{1}}+R^{anti-parallel, 
\rho_{1}\sigma_
{1}}_{
Free-Bound,\mu_{1}\nu_{1}}
\nonumber\\
&& \ \ \ \ \ \ \ \ \ \ \ \ \ \  \ \  \ \  \ \
+
R^{parallel, \rho_{1}\sigma_{1}}_{Free-Free,\mu_{
1}\nu_{1}}
+R^{parallel, \rho_{1}\sigma_{1}}_{Bound-Bound,\mu_{1}\nu_{1}}+R^{parallel, 
\rho_{1}\sigma_{
1}}_{Free-
Bound,\mu_{1}\nu_{1}}\Big]
\Bigg\}.
\label{T2}
\end{eqnarray}
Finally, using   (\ref{T1}) and (\ref{T2}) we can easily extract the explicit form for 
the number of degrees of freedom   of the $M1$-brane  on the surface and in the bulk  in 
terms of curvatures, namely
\begin{eqnarray}
&&\!\!\!\!\!\!\!\!\!\!\!\!\!
   N_{sur}^{M1}=- \int d^{2}x \sqrt{-g}
  \Biggl\{ (1-m_{g}^{2})
 \Big[( R_{Free-Free}^{parallel})^{2}+( R_{Free-Bound}^{parallel})^{2}+( 
R_{Bound-Bound}^{
parallel})^{2}
\nonumber
\\&& 
\ \ \ \ \ \ \ \ \ \ \ \ \ \ \ \ \ \ \ \ \ \ \ \ \ \  \ \ \ \ \ \  \  \ \ \ \ 
+(R_{Free-Free}^{parallel} 
R_{Free-Free}^{anti-parallel})\partial^{2}(
R_{Free-Free}^{parallel} )
\nonumber
\\&&
\ \ \ \ \ \ \ \ \ \ \ \ \ \ \ \ \ \ \ \ \ \ \ \ \ \  \ \ \ \ \ \  \  \ \ \ \ 
+(R_{Free-Bound}^{parallel} 
R_{Free-Bound}^{anti-parallel})\partial^{2}(R_
{Free-Bound}^{parallel} )
\nonumber
\\&& 
\ \ \ \ \ \ \ \ \ \ \ \ \ \ \ \ \ \ \ \ \ \ \ \ \ \  \ \ \ \ \ \  \  \ \ \ \ 
+
(R_{Bound-Bound}^{parallel} 
R_{Bound-Bound}^{anti-parallel})
\partial^{2}(R_{Bound-Bound}^{parallel} )
\Big]
\nonumber\\&&
\ \ \ \ \ \ \ \,   
+      m_{g}^{2}\lambda^{2}\delta_{\rho_{1}\sigma_{1}}^{\mu_{1}\nu_{1}}           
\Big[R^{parallel, \rho_{1}\sigma_{1}}_{Free-Free,\mu_{1}\nu_{1}}+R^{parallel, \rho_{1}
\sigma_ { 1 }
 }_{Bound-Bound,\mu_{1}\nu_{1}}+              
R^{parallel, \rho_{1}\sigma_{1}}_{Free-Bound,\mu_{1}\nu_{1}}\Big]
\Biggr\},
\label{T3}
\end{eqnarray}
\begin{eqnarray}
&& 
\!\!\!\!\!\!\!\!\!\!\!\!\!\!\!
N_{bulk}^{M1}=\int d^{2}x \sqrt{-g}
  \Biggl\{(1-m_{g}^{2})
   \Big[( R_{Free-Free}^{anti-parallel})^{2}+( 
R_{Free-Bound}^{anti-parallel})^{2}+( R_{
Bound-Bound}^{anti-parallel})^{2}
\nonumber \\
&& \ \ \ \ \ \ \ \ \ \ \ \ \ \ \ \ \ \ \ \ \ \ \ \ \ \  \ \ \ \  \  \  \ \ 
+(R_{Free-Free}^{parallel} 
R_{Free-Free}^{anti-
parallel})\partial^{2}(R_{Free-Free}^{anti-parallel} )
\nonumber \\
&& \ \ \ \ \ \ \ \ \ \ \ \ \ \ \ \ \ \ \ \ \ \ \ \ \ \  \ \ \ \ \ \   \ \ 
+(R_{Free-Bound}^{parallel} 
R_{Free-Bound}^{
anti-parallel})\partial^{2}(R_{Free-Bound}^{anti-parallel} )
\nonumber 
\\&& 
\ \ \ \ \ \ \ \ \ \ \ \ \ \ \ \ \ \ \ \ \ \ \ \ \ \  \ \ \ \ \ \  \  \ 
+
(R_{Bound-Bound}^{
parallel} R_{Bound-Bound}^{anti-parallel})\partial^{2}(R_{Bound-Bound}^{anti-parallel} 
)\Big]
\nonumber
\\&&
   \ \ \  \ \,  + m_{g}^{2}\lambda^{2}\delta_{\rho_{1}\sigma_{1}}^{\mu_{1}\nu_{1}}
  \Big[R^{anti-parallel, \rho_{1}\sigma_{1}}_{Free-Free,\mu_{1}\nu_{1}}+
  R^{anti-parallel, \rho_{1}\sigma_{1}}_{Bound-Bound,\mu_{1}\nu_{1}}+           
 R^{anti-parallel, \rho_{1}\sigma_{1}}_{Free-Bound,\mu_{1}\nu_{1}}  \Big]
\Biggr\}.
\label{T4}
 \end{eqnarray}
These expressions imply that the number of degrees of freedom on the surface depends 
only on the curvature of parallel spins, and that the number of degrees of freedom 
in the bulk depends only on the curvature of anti-parallel spins. On the other 
hand, as it was shown in  (\ref{Rl1l2a}) and (\ref{Rl1l2b}), the 
curvature of parallel spins depends on the brane length and the curvature of 
anti-parallel spins depends on the separation distance between branes. Thus, if the 
branes approach mutually, where as it was shown  in the previous subsection  the 
curvature 
of parallel spins increase and curvature of anti-parallel spins decreases, then  the 
number of degrees of freedom on the surface increases, while the number of 
degrees of freedom in the bulk decreases.  

Finally, applying the expressions (\ref{T3}), (\ref{T4}) in the explicit example of the 
previous subsection, namely using (\ref{Rl1l2a}),(\ref{Rl1l2b}), we obtain:
\begin{eqnarray}
&&
\!\!\!\!\!\!\!\!\!\!\!\!\!\!\!\!\!\!\!\!\!\!
N_{bulk}^{M1}(t)
\approx
  e^{2m_{g}^{2}\lambda^{2}(1-\frac{t}{t_{s}}+m_{g}^{2})}\,
\Biggl\{
 \frac{1}{6m_{g}^{2}\lambda^{2} t_{s}} 
\left(1-\frac{t}{t_{s}}+m_{g}^{2}\right)^{6}
+\frac{1}{5t_{s}}
\left(1-
\frac{t}{t_{s}}+m_{g}^{2}\right)^{5}
\nonumber\\
&&    \ \ \ \ \  \ \ \ \ \  \ \ \ \ \     \ \ \ \ \  \ \ \ \ \ \ \
+
\frac{2}{5t_{s}}
\left(1-\frac{t}{t_{s}}+m_{g}^{2}\right)^{3}
+
\frac{1}{2t_{s}}
\left(1-\frac{t}{t_{s}}+    
m_{g}^{2}\right)^{2}\Biggr\},
\label{T3bb}
\end{eqnarray}
and 
\begin{eqnarray}
&&
\!\!\!\!\!\!\!\!\!\!\!\!\!\!\!\!\!\!\!\!
N_{sur}^{M1}(t) \approx
  e^{-2m_{g}^{2}\lambda^{2}(1-\frac{t}{t_{s}}+m_{g}^{2})}
\Biggl\{       
\frac{1}{3t_{s}m_{g}^{2}\lambda^{2}}\left(1-\frac{t}{t_{s}}+m_{g}^{2}\right)^{-3}+\frac{1}
{4t_{s}      
}\left(1-\frac{t}{t_{s}}+m_{g}^{2}\right)^{-4}
\nonumber\\
&&    \ \ \ \ \  \ \ \ \ \  \ \ \ \ \  \ \ \ \ \  \ \ \ \ \ \ \ \ \
+
\frac{1}{2t_{s}}\left(1-\frac{t}{t_{s}}+m_{g}^{2}\right)^{-2}+
\frac{1}{t_{s}}\left(1-\frac{t}{t_{s}}+m_{g}^{2}\right)^{-1}\Biggr\}.
\label{T4bb}
\end{eqnarray}
Let us now use the results  (\ref{T3bb}), (\ref{T4bb}) in order to calculate the total 
number of degrees of freedom for the $Mp$-branes. Straightforwardly we find:
\begin{eqnarray}
&& 
\!\!\!\!\!\!\!\!\!\!\!\!\!\!
N_{bulk}^{Mp}(t) =
\sum_{n=1}^{p}   
\delta_{a_{1},a_{2}...a_{n}}^{b_{1}b_{2}....b_{n}}
\left(N_{bulk}^{M1}\delta_{b_1}^{a_1}\right)
\cdots\left(N_{bulk}^{M1}\delta_{b_n}^{a_n}\right)
 \nonumber
 \\
 &&  \ \ \ \
  \approx
  e^{2m_{g}^{2}\lambda^{2}(1-\frac{t}{t_{s}}+m_{g}^{2})}\,
\sum_{n=1}^{p}
\Biggl\{
\left[\frac{1}{m_{g}^{2}\lambda^{2} t_{s}(
n+5)}\right]
\left(1-\frac{t}{t_{s}}+m_{g}^{2}\right)^{n+5}\nonumber\\
&&   \ \ \ \ \  \ \ \ \ \  \ \ \ \ \  \ \ \
+\left[\frac{1}{t_{s}(n+4)}
\right]
\left(1-
\frac{t}{t_{s}}+m_{g}^{2}\right)^{n+4}+
\left[\frac{1}{t_{s}(\frac{n}{2}+2)}\right]
\left(1-\frac{t}{t_{s}}+m_{g}^{2}\right)^{n+2}
\nonumber\\
&&    \ \ \ \ \  \ \ \ \ \  \ \ \ \ \  \ \ \
+\left[
\frac{1}{t_{s}(n+1)}\right]
\left(1-\frac{t}{t_{s}}+    
m_{g}^{2}\right)^{n+1}\Biggr\},
\label{T5aa}
\end{eqnarray}
and   
\begin{eqnarray}
   &&\!\!\!\!\!\!\!\!\!\!\!\!\!\!\!\!\!
N_{sur}^{Mp}(t) =\sum_{n=1}^{p}   
\delta_{a_{1},a_{2}...a_{n}}^{b_{1}b_{2}....b_{n}}
\left(N_{sur}^{M1}\delta_{b_1}^{a_1}\right)
\cdots\left(N_{sur}^{M1}\delta_{b_n}^{a_n}\right)
\nonumber\\
&&\!\!\!\!\!
 \approx
  e^{-2m_{g}^{2}\lambda^{2}(1-\frac{t}{t_{s}}+m_{g}^{2})}\,
\sum_{n=1}^{p}
\Biggl\{\frac{1}{t_{s}(n+1)(1-\frac{t}{t_{s}}+m_{g}^{2})^{
n+1}}+\frac{1}{t_{s}(n+3)      
(1-\frac{t}{t_{s}}+m_{g}^{2})^{n+3}}
\nonumber\\
&&    \ \ \ \ \  \ \ \ \ \  \ \ \ \ \  \ \ \   \ \ \ \ \  \ \
+
\frac{1}{t_{s}(n)(1-\frac{t}{t_{s}}+m_{g
}^{
2})^{n}}+      
\frac{1}{t_{s}m_{g}^{2}\lambda^{2}(n+2)(1-\frac{t}{t_{s}}+m_{g}^{2})^{n+2}}\Biggr\}.
\label{T5}
\end{eqnarray}

As we observe, and as expected, the qualitative result of the individual $M1$brane is 
transfered to the $Mp$-brane too, namely as time passes, and due to the presence of the 
exponentials in   (\ref{T5aa}), (\ref{T5}), 
by mutual approach of the branes the number of
degrees of freedom on the surface increases, while the number of degrees of freedom in 
the bulk decreases. Hence, during the approach of branes and anti-branes, a large amount 
of energy is dissolved in branes, and their length grows. Due to 
(\ref{Rl1l2a}), this length increase leads to the increase of the curvature of 
parallel spins, and then, by increasing the curvature of parallel spins, the
number of degrees of freedom on their surface increases (see relation (\ref{T3})). 
On the other hand, by decreasing the separation between branes, the curvature of 
anti-parallel spins decreases (see relation (\ref{Rl1l2b})) 
which leads to a decrease in number of degrees of freedom in the bulk (see (\ref{T4})). 
In summary, these lead to the emergence of an inequality between the number of degrees of 
freedom,  and hence according to Padmanabhan approach the universe  
expands.

\section{The universe contraction from the Pauli exclusion 
principle}
\label{Sectioncontraction}

In the previous sections,  we saw that by mutual approach of the branes the number of
degrees of freedom on the surface increases, while the number of degrees of freedom in 
the bulk decreases, and due to this inequality between the number of degrees of freedom 
in the surface and in the bulk, the universe expands. It would be interesting to see 
whether one could have a scenario behaving in the opposite way, i.e resulting to a 
contracting universe. Such a construction is performed in the present section.

As we saw in solutions (\ref{s22a}),(\ref{s22}), as time passes $l_1$ decreases and $l_2$ 
increases. Hence, near the colliding moment at $t=t_{s}$ the quantity under the square 
root in action (\ref{w13}) becomes negative, i.e. the square energy of the system becomes 
negative and tachyonic states may be created 
\cite{Sepehri:2016jfx,Sepehri:2015eea,Sepehri:2016sjq,Sepehri:2016svv, Sepehri:2015ara}. 
In order to handle this severe problem we introduce the mechanism of 
\cite{Sepehri:2016jfx,Sepehri:2015eea,Sepehri:2016sjq,Sepehri:2016svv,Sepehri:2015ara},
 which allows branes to compactify on a circle, and attractive gravity to change 
to repulsive gravity, which forbids the branes to come closer than a minimum distance and 
thus the energy square to remain always positive.

In particular, we define ${\displaystyle <\psi^{3}>=\frac{r}{l_{s}^{3/2}}}
$ with $l_{s}$ the string coupling, and where  $r$ is the radius of a circle that branes 
are compactified on it.    Using this choice, andas we show in Appendix 
\ref{compactified} in relation   (\ref{sss24}), we acquire
\begin{eqnarray}
&& 
\Sigma_{a,b,c=0}^{3} \langle
F^{abc},F_{abc}\rangle 
 = 
6\left(\frac{r^{2}}{l_{s}^{3}}\right)\Sigma_{a,b=0}^{2}\partial_{U}^{a}\Psi_{L,b}\partial_
{a,U}\Psi_{L}^{b}
+6\left(\frac{r^{2}}{l_{s}^{3}}\right)\Sigma_{a,b=0}^{2}\partial_{L}^{a}\Psi_{U,b}
\partial_{a,L}\Psi_{U}^{b}\nonumber \\
&&
-12\left(\frac{r^{2}}{l_{s}^{3}}\right)\Sigma_{a,b=0}^{2}\partial_{U}^{a}\Psi_{L,b}
\partial_{a,L}\Psi_{U}^{b}
 +E_{Extra} ,
 \label{s24}
\end{eqnarray}
where $ E_{Extra}=- 
6\Sigma_{a,b,c,a',b',c'=0,\neq3}^{2}\varepsilon_{abcD}\varepsilon_{a'b'c'}^{D}X^{a}X^{b}
\psi_{L}^{
b}\psi_{L}^{b'} $.
Expression (\ref{s24}) indicates  that by compacting branes the sign of gravity changes 
and repulsive gravity emerges. In fact, the symmetry of the system is broken again and 
terms related to attractive force are canceled. This reveals that the repulsive 
gravity between parallel spins is the underlying reason for the appearance of Pauli 
exclusion principle. 

As we show in (\ref{s25}), the curvature (\ref{w7new}) becomes:
\begin{eqnarray}
  && R_{MN}=\langle
 F^{\rho}\smallskip_{\sigma\lambda},F^{\lambda}\smallskip_{\mu\nu}\rangle= 
6\left(\frac{r^{2}}{l_{s}^{3}}\right)\Sigma_{\rho,\mu=0}^{2}\partial_{U}^{\rho}\psi_{L,
\sigma}\partial_{U,\mu}\psi_{L,\nu}+
 6\left(\frac{r^{2}}{l_{s}^{3}}\right)\Sigma_{\rho,\mu=0}^{2}\partial_{L}^{\rho}\Psi_{U,
\sigma}\partial_{L,\mu}\Psi_{U,\nu}
\nonumber\\&&
-
  12\left(\frac{r^{2}}{l_{s}^{3}}\right)\Sigma_{\rho,\mu=0}^{2}\partial_{L}^{\rho}\Psi_{U,
\sigma}\partial_{U,\mu}\Psi_{L,\nu}=
   R_{MN,parallel}-R_{MN,anti-parallel}.
\label{s25}
 \end{eqnarray}
 Hence, following the calculations of Appendix \ref{compactified}, in relation 
(\ref{ss30bc}) we extract the total compactified $Mp$-branes action as:
 \begin{eqnarray}
&& 
\!\!\!\!\!\!\!\!
S_{Mp} = - \!
 \int 
d^{p+1}x\sqrt{-g}
\Biggl\{
\Biggl\{\sum_{n=1}^{p}
\Big(1-m_{g}^{2})^n
\Big[(R^{parallel}_{Free-Free}+R^{parallel}_{
Free-Bound}
+R^{parallel}_{Bound-Bound})
\nonumber\\
&&
\ \ \ \ \ \ \ \ \ \ \ \ \ \ \ \ \ \ \ \ \ \ \ \ \ \ \ \ \ \ \ \ \ \ \ \ \ \ \   \ \ \ \ \ 
 \ \ \ \
- 
(R_{Free-Free}^{anti-parallel}+R^{anti-parallel}_{Free-Bound}+R^{anti-parallel}_{
Bound-Bound})\Big]^{n}
 \nonumber\\
 &&  \ \ \ \ \  
 +
\sum_{n=1}^{p}
m_{g}^{2n}\lambda^{2n}\delta^{\rho_{1}\sigma_{1}...\rho_{n}\sigma_{n}}_{\mu_
{1}\nu_{
1}...\mu_{n}\nu_{n}} 
\Big[
\Big(R^{parallel, \rho_{1}\sigma_{1}}_{Free-Free,\mu_{1}\nu_{1}}+R^{parallel, \rho_{1}
\sigma_ {1}}_{ Bound-Bound ,
\mu_{1}\nu_{1}}+
 R^{parallel, \rho_{1}\sigma_{1}}_{Free-Bound,\mu_{1}\nu_{1}}\Big)
 \nonumber\\
 && 
 \ \ \ \ \ \ \ \ \ \ \ \ \ \ \ \ \ \ \ \ \ \ \  \ \ \ \  \ \ \ \ \ \ \
 -
\Big(
R^{anti-parallel, \rho_{1}\sigma_{1}}_{Free-Free,\mu_{1}\nu_{1}}+R^{anti-parallel, 
\rho_{1}
\sigma_{1}}_{ Bound-Bound,
\mu_{1}\nu_{1}}+ 
R^{anti-parallel, \rho_{1}\sigma_{1}}_{Free-Bound,\mu_{1}\nu_{1}}\Big)\Big]
\nonumber\\
&&  \ \ \ \ \ \ \ \   \ \ \ 
\cdots 
\times
 \Big[
 \Big(
 R^{parallel, \rho_{n}\sigma_{n}}_{Free-Free,\mu_{n}\nu_{n}}+ 
R^{parallel, \rho_{n}\sigma_{n}}_{Bound-Bound,\mu_{n}\nu_{n}}+R^{parallel, 
\rho_{n}\sigma_{n
} }_{ Free-Bound,
\mu_{n}\nu_{
n}}\Big)
\nonumber\\
&&  \ \ \ \ \ \ \ \ \ \ \  \ \ \ \ \ \ \ \
-
\Big(
R^{anti-parallel, \rho_{n}\sigma_{n}}_{Free-Free,\mu_{n}\nu_{n}}+R^{anti-parallel, 
\rho_{n}
\sigma_{n}}_{ Bound-Bound,
\mu_{n}\nu_{n}}+ 
R^{anti-parallel, \rho_{n}\sigma_{n}}_{Free-Bound,\mu_{n}\nu_{n}}\Big)
\Big]
\Biggr\}^{\frac{1}{2}}
\Biggr\}.
\label{s30}
 \end{eqnarray}
Hence, from the above action we deduce that for a system of $Mp$-anti-$Mp$-branes, near  
the colliding moment ($t=t_{s}$) the shape of action changes and the sign of gravity 
reverses, which prevents the collision and  forbids the branes to come closer than a 
minimum distance and thus the energy square to remain always positive.

We proceed by calculating the couplings of parallel and anti-parallel spins, using the 
method of the previous section. We consider for simplicity the choices 
(\ref{choicepsi1new}),(\ref{choicepsi2new}). Substituting  these into (\ref{s25})
we obtain 
\begin{eqnarray}      
&&R_{Free-Free}^{parallel}=R_{Free-Bound}^{parallel}=R_{Bound-Bound}^{parallel}\approx 
\frac{(l'_
{
2})}{(l_{2})}
\label{Rl1l2asec3} 
\\&&R_{Free-Free}^{anti-parallel}=R_{Free-Bound}^{anti-parallel}=R_{Bound-Bound}^{
anti-parallel}\approx \frac{(l'_
{
1})}{(l_{1})},
\label{Rl1l2bsec3}
\end{eqnarray}
and therefore inserting into  action (\ref{s30}) we finally acquire:
\begin{eqnarray}
 &&  S_{Mp-brane}\approx V\int  dt\, \Sigma_{n=1}^{p} 
\left\{
12\left[m_{g}^{2}\lambda^{2}+(1-m_{g}
^{2})\right]\left[\frac{(l'_{2})^{2}}{(l_{2})^{2}}-
\frac{(l'_{1})^{2}}{(l_{1})^{2}}\right]
 \right\}^{\frac{n}{2}},
 \label{s31}
\end{eqnarray}
where $'$ denotes the derivative respect to time. 

Variation of action  (\ref{s31}) leads to the following equations of motion:
\begin{eqnarray}        
&&\!\!\!\!\!\!\!\!\!\!\!\!\!\!\!\!\!\!\!\!
\Sigma_{n=1}^{p}
\Biggl\{ \left[m_{g}^{2}\lambda^{2}+(1-m_{g}^{2})\right]\left[\frac{(l'_
{
2})}{(l_{2})^{2}}\right] 
\left\{ \left[m_{g}^{2}\lambda^{2}+(1-m_{g}^{2})\right]\left[\frac{(l'_{2})^{2}}{(l_{2})^{
2}}-\frac{(l'_{1})^{2}}
        {(l_{1})^{2}}\right]\right\}^{\frac{n}{2}-1}\Biggr\}'    
 \nonumber\\
&&\!\!\!\!\!\!\!\!\!\!
=\Sigma_{n=1}^{p}
\Biggl\{ \left[m_{g}^{2}\lambda^{2}+(1-m_{g}^{2})\right]\left[\frac{(l'_{
2}
)^{2}}{(l_{2})^{3}}\right] 
\left\{ \left[m_{g}^{2}\lambda^{2}+(1-m_{g}^{2})\right]
\left[\frac{(l'_{2})^{2}}{(l_{2})^{2}}-\frac{(l'_{1})^{2}}{(l_{1})^{2}}\right]
   \right\}^{\frac{n}{2}-1}\Biggr\},
   \label{s32}
 \end{eqnarray}
and
\begin{eqnarray}          
&&\!\!\!\!\!\!\!\!\!\!\!\!\!\!\!\!\!\!\!\!
\Sigma_{n=1}^{p}
\Biggl\{\left[m_{g}^{2}\lambda^{2}+(1-m_{g}^{2})\right]\left[\frac{(l'_
{1})}{(l_{1})^{2}}\right] 
\left\{\left[m_{g}^{2}\lambda^{2}+(1-m_{g}^{2})\right]\left[\frac{(l'_{2})^{2}}{(l_{2})^
{2}}-\frac{(l'_{1})^{2}}{(l_{1})^{2}}\right]
        \right\}^{\frac{n}{2}-1}\Biggr\}'
          \nonumber\\
&&\!\!\!\!\!\!\!\!\!\!
 =          
\Sigma_{n=1}^{p}
\Biggl\{
\left[m_{g}^{2}\lambda^{2}+(1-m_{g}^{2})\right]\left[\frac{(l'_{
1})^{2}}{(l_{1})^{3}}\right] 
\left\{[m_{g}^{2}\lambda^{2}+(1-m_{g}^{2})]\left[\frac{(
l'_{2})^{2}}{(l_{2})^{2}}-\frac{(l'_{1})^{2}}{(l_{1})^{2}}\right]\right\}^{\frac{n}{2}-1}
  \Biggr\}.
\label{s33}
\end{eqnarray}
The solutions of these equations are approximately:
\begin{eqnarray}
  &&l_{1}(t)
  \approx 
\Sigma_{n=1}^{p}
\left[e^{m_{g}^{2}\lambda^{2}(\frac{t}{t_{
s}}-1+m_{g}^{2})^{n+2}}-1\right]
\nonumber\\
&&
l_{2}(t)
\approx 
\Sigma_{n=1}^{p}\left(\frac{t}{
t_{s}}-1+m_{g}^{2}\right)^{-n}\left[1+e^{m_{g}^{2}\lambda^{2}(\frac{t}{t_{
s}}-1+m_{g}^{2})^{-1}}\right].
\label{s34}
\end{eqnarray}
 These solutions reveal that as time passes the coupling between anti-parallel 
spins, i.e. $l_1(t)$, increases and tends to infinity near the colliding point $t_s$, 
while the coupling between parallel spins, i.e.  $l_2(t)$, decreases and tends to zero 
near  $t_s$. Hence, the universe, which lies on an $M3$-brane which has been constructed 
from $M1$-branes, contracts.

Let us now calculate the number of degrees of freedom  in the bulk and on the surface 
for an $M1$-brane, similarly to the previous section. The sum over the number of degrees 
of freedoms is in direct relation with the energy of the $M1$-brane, namely 
\cite{Sepehri:2016jfx}:
\begin{eqnarray}
&& 
\!\!\!\!\!\!\! 
N_{sur}^{M1}+N_{bulk}^{M1}\approx E_{M1}
\nonumber \\
&& \ \ \  \ \ \  \ \ \  \ \  \ \  \ \,
=
- \int d^{2}x \sqrt{-g}
\Biggl\{
(1-m_{g}^{2})
\Big[R^{parallel}_{Free-Free}+R^{parallel}_{
Free-Bound}+R^{parallel}_{ Bound-
Bound}
\nonumber\\
&&\ \ \  \ \ \ \ \ \  \ \ \ \ \ \  \ \ \ \ \ \  \ \ \ \ \ \  \ \ \ \ \ \  \ \ \ \ \ \  \ 
\ 
\  \ \  \ \  \ \  \ \ \,
-
R^{anti-parallel}_{Free-Free}-R^{anti-parallel}_{Free-Bound}-R^{anti-parallel}_{
Bound-Bound}\Big]
\nonumber\\
&&\ \ \  \ \ \ \ \ \  \ \   \ \ \ \  \ \ \ \ \ \  \ \ \ \ 
+            
m_{g}^{2}\lambda^{2}\delta_{\rho\sigma}^{\mu\nu}
\Big[R^{parallel, \rho\sigma}_{Free-Free,\mu\nu}+R^{ parallel,
\rho\sigma}_{Bound-Bound,\mu\nu}+                 
R^{parallel, \rho\sigma}_{Free-Bound,\mu\nu}
\nonumber\\
&&\ \ \  \ \ \ \ \ \  \ \ \ \ \ \  \ \ \ \ \ \  \ \ \ \  \  \   \ \ \  \ 
 -
R^{anti-parallel, \rho\sigma}_{Free-
Free,\mu\nu}-R^{anti-parallel, \rho\sigma}_{Bound-Bound,\mu\nu}-                 
R^{anti-parallel, \rho\sigma}_{Free-Bound,\mu\nu}\Big]   
\Biggr\}
\label{s35}.
\end{eqnarray}               
On the other hand, since the difference between the number of degrees of freedom in the 
bulk and on the surface causes that system energy to change with time 
\cite{Sepehri:2014jla,Sepehri:2016jfx,Sepehri:2016sjq,Sepehri:2016svv}, and 
assume that all  curvatures of free 
and bound fermions have the same behavior, we can calculate:
 \begin{eqnarray}   
  &&
  \!\!\!\!\!\!\! 
N_{sur}^{M1}-N_{bulk}^{M1}\approx
\frac{\partial E_{system}}{\partial t}=\frac{\partial E_{system}}{\partial 
l_{1}}\frac{\partial l_{1}}{\partial 
t}+\frac{\partial E_{system}}{\partial l_{2}}\frac{\partial l_{2}}{\partial t}
\nonumber 
\\&&
 \ \ \  \ \ \  \ \ \  \ \  \ \  \ \,
    =  -  \int d^{2}x \sqrt{-g}
\Biggl\{
(1-m_{g}^{2})\Big[R^{parallel}_{Free-Free}+R^{parallel}_{
Free-Bound}
+R^{parallel}_{Bound-Bound}
\nonumber\\
&&\ \ \  \ \ \ \ \ \  \ \ \ \ \ \  \ \ \ \ \ \  \ \ \ \ \ \  \ \ \ \ \ \  \ \ \ \ \ \  \ 
\ 
\  \ \  \ \  \ \  \ \ \,
+
R^{anti-parallel}_{Free-Free}+R^{anti-parallel}_{Free-Bound}+R^{anti-parallel}_{
Bound-Bound}\Big]
\nonumber\\
&&\ \ \  \ \ \ \ \ \  \ \   \ \ \ \  \ \ \ \ \ \  \ \ \ \  
+            
m_{g}^{2}\lambda^{2}\delta_{\rho\sigma}^{\mu\nu}
\Big[R^{parallel, \rho\sigma}_{Free-Free,
\mu\nu}+R^{ parallel,
\rho\sigma}_{Bound-Bound,\mu\nu}+ R^{parallel, \rho\sigma}_{Free-Bound,\mu\nu})  
\nonumber\\
&&\ \ \  \ \ \ \ \ \  \ \ \ \ \ \  \ \ \ \ \ \  \ \ \ \  \  \   \ \ \  \
+
R^{anti-parallel, \rho\sigma}_{Free-
Free,\mu\nu}+ 
R^{anti-parallel, \rho\sigma}_{Bound-Bound,\mu\nu}+R^{anti-parallel, 
\rho\sigma}_{Free-Bound , \mu\nu}
\Big]
\Biggr\}.
\label{s36}
\end{eqnarray}
Hence, using (\ref{s35}) and (\ref{s36}) we can calculate the explicit form of the number 
of degrees of freedom on the surface and in the bulk in terms of curvatures as
\begin{eqnarray}
   && 
   \!\!\!\!\!\!\!\!\!\!\!\!\!\!\!\!
   N_{sur}^{M1}=- \int d^{2}x 
   \sqrt{-g}
\Biggl\{
(1-m_{g}^{2})
\Big[R^{parallel}_{Free-Free}+R^{parallel}_{Free-Bound}+R^{parallel}_{Bound-Bound}
\Big]
\nonumber\\
&& \ \ \ \ \ \ \ \ \ \ \ \ \ \  
+                                                   
m_{g}^{2}\lambda^{2}\delta_{\rho\sigma}^{\mu\nu}
\Big[R^{parallel, \rho\sigma}_{Free-Free,\mu\nu}+R^{parallel,
\rho\sigma}_{Bound-Bound,\mu\nu}+                 
R^{parallel, \rho\sigma}_{Free-Bound,\mu\nu}\Big] 
\Biggr\},
\label{s37}
  \end{eqnarray}
  and
   \begin{eqnarray}
   &&
      \!\!\!\!\!\!\!\!\!\!\!\!\!\!\!\!
      N_{bulk}^{M1}=- \int d^{2}x  \sqrt{-g}               
 \Biggl\{(1-m_{g}^{2})
 \Big[
R^{anti-parallel}_{Free-Free}+R^{anti-parallel}_{Free-Bound}+R^{anti-parallel}_{
Bound-Bound}\Big]
\nonumber\\
&& \ \ \ \ \ \ \ \ \ \ \ \ 
+                               
m_{g}^{2}\lambda^{2}\delta_{\rho\sigma}^{\mu\nu}
\Big[ R^{anti-parallel, \rho\sigma}_{Free-Free,\mu\nu}+R^{anti-parallel,
\rho\sigma}_{Bound-Bound,\mu\nu}+                 
R^{anti-parallel, \rho\sigma}_{Free-Bound,\mu\nu}
\Big]
\Biggr\}.
\label{s38}
\end{eqnarray}
These expressions show that the number of degrees of freedom on the surface depends on 
the curvature of parallel spins, and the number of degrees of freedom in a bulk has a 
direct relation with curvature of anti-parallel spins.

Finally, applying the expressions (\ref{s37}), (\ref{s38}) in the explicit example of   
(\ref{Rl1l2asec3}),(\ref{Rl1l2bsec3}), with (\ref{s34}),  we 
obtain:
\begin{eqnarray}
&&
\!\!\!\!\!\!\!\!\!\!\!\!\!\!\!\!\!\!\!\!
N_{bulk}^{M1}(t) \approx
  \left(\frac{m_{g}^{2}\lambda^{2}}{t_{s}}\right)^{
  2}\left(\frac{t}{t_{s}}-1+m_{g}^{2}\right)^{2},
\label{T4bbcc}
\end{eqnarray}
and 
\begin{eqnarray}
&&
\!\!\!\!\!\!\!\!\!\!\!\!\!\!\!\!\!\!\!\!\!\!
N_{sur}^{M1}(t)
\approx
  \left[\frac{1}{t_{s}(\frac{t}{t_{s}}-1+m_{g}^{2})^{
  2}}\right]
  \Biggl\{
  1+\left(\frac{m_{g}^{2}\lambda^{2}}{t_{s}}\right)       
  \frac{1}{1+e^{
  \left[-\frac{m_{g}^{2}\lambda^{2}}
  {(\frac{t}{t_{s}}-1+m_{g}^{2})}\right]}}\Biggr\}.
\label{T3bbcc}
\end{eqnarray}
Let us now use the results  (\ref{T4bbcc}), (\ref{T3bbcc}) in order to calculate the 
total number of degrees of freedom for the $Mp$-branes by 
summing over number of degrees of freedom of $M1$-branes. Straightforwardly we find:
 \begin{eqnarray}
     && 
\!\!\!\!\!\!\!\!\!\!\!\!\!\!
N_{bulk}^{Mp}(t) =
\sum_{n=1}^{p}   
\delta_{a_{1},a_{2}...a_{n}}^{b_{1}b_{2}....b_{n}}
\left(N_{bulk}^{M1}\delta_{b_1}^{a_1}\right)
\cdots\left(N_{bulk}^{M1}\delta_{b_n}^{a_n}\right)
 \nonumber
 \\
 &&  \ \ \ \
  \approx 
\sum_{n=1}^{p}\left(\frac{m_{g}^{2}\lambda^{2}}{t_{s}}\right)^{
n+1}\left(\frac{t}{t_{s}}-1+m_{g}^{2}\right)^{n+1},
\label{s39}
  \end{eqnarray}
  and
   \begin{eqnarray}
&&\!\!\!\!\!\!\!\!\!\!\!\!\!\!\!\!\!
N_{sur}^{Mp}(t) =\sum_{n=1}^{p}   
\delta_{a_{1},a_{2}...a_{n}}^{b_{1}b_{2}....b_{n}}
\left(N_{sur}^{M1}\delta_{b_1}^{a_1}\right)
\cdots\left(N_{sur}^{M1}\delta_{b_n}^{a_n}\right)
\nonumber\\
&&\!\!\!\!\!
 \approx
\sum_{n=1}^{p}
\left[\frac{1}{t_{s}^{n}(\frac{t}{t_{s}}-1+m_{g}^{2})^{
2n}}\right]
\Biggl\{
1+\left(\frac{m_{g}^{2}\lambda^{2}}{t_{s}}\right)       
\frac{1}{1+e^{
\left[-\frac{m_{g}^{2}\lambda^{2}}
{(\frac{t}{t_{s}}-1+m_{g}^{2})}\right]}}\Biggr\}^{n}.
\label{s40}
\end{eqnarray}
             
As we can see from these expressions, the qualitative result of the individual $M1$brane 
is transfered to the $Mp$-brane too, namely as time passes the number of
degrees of freedom on the surface decreases, while the number of degrees of freedom in 
the bulk increases. We stress that this behavior is exactly the opposite of the one in 
the previous section. Hence, by compacting the branes repulsive gravity emerges 
and branes tend to mutual remove, and thus through Padmanabhan approach
this leads to to a contracting universe. 
 
We close this section by mentioning that since in our present construction we can obtain 
both expansion and contraction, one could build more complicated models in which bouncing 
or cyclic evolution \cite{Novello:2008ra,Qiu:2013eoa} could also emerge. Such an 
extensive analysis lies beyond the scope of the present work and it is the matter for a future 
investigation.

\section{Conclusions} 
\label{Conclusions}

In this work we have shown that the Pauli exclusion principle in a system of $M0$-branes 
can result to the expansion and contraction of the universe which is located on an 
$M3$-brane. In particular, initially we consider that there exist only scalar fields that 
play the role of graviton scalar modes, and are attached to $M0$-branes. At this stage, 
the system has a high symmetry and no gauge fields or fermions live on the $M0$-branes. 
Then, $M0$-branes join mutually and form pairs of $M1$-anti-$M1$-branes. In this new 
stage, the symmetry of the system is broken, and gauge fields that live on the $M1$-branes 
arise and play the role of graviton tensor modes. Therefore, the attractive force between 
the $M1$ and anti-$M1$ branes leads to breaking of the symmetry of $M1$-branes, and thus 
the lower and upper parts of thee $M1$-branes are no longer the same. Consequently, the 
gauge fields that live on the $M1$-branes, and the scalar fields which are attached 
symmetrically to all parts of these branes, decay to fermions that attach 
anti-symmetrically to the upper and lower parts of the branes. Since fermions 
that live on the upper part of $M1$-branes have the opposite spin with respect to 
fermions that are placed on the lower part of the $M1$-branes, we deduce that when 
upper and lower spins are mutually linked $M1$-branes emerge and this implies the 
appearance of an attractive force between anti-parallel spins. Finally, by joining 
$M1$-branes, $M3$-branes are created, and our universe is located on one of them.
 
Since there is no link between parallel spins in $M1$-branes, they cannot construct a 
stable system, and hence the Pauli exclusion principle emerges. By closing $M1$-branes 
mutually, the curvatures produced by parallel spins will be different from the curvatures 
produced by anti-parallel spins, and this leads to an inequality between the number of 
degrees of freedom on the boundary surface and the number of degrees of freedom in the 
bulk region. This behavior is inherited in the  $M3$-brane on which the universe is 
located, and hence according to Padmanabhan approach this leads to the emergence of the 
universe  expansion and contraction. 

In our analysis we provided two specific examples, one that leads to universe expansion 
and one that leads to contraction. Hence, it becomes clear that one could construct more 
complicated models in which bouncing or cyclic evolution could also emerge. In summary, 
the universe evolution is a result of the difference between curvatures of parallel and 
anti-parallel spins and of the Pauli exclusion principle.

\begin{acknowledgments}
The work by A. Sepehri has been financially supported by the Research Institute 
for Astronomy and Astrophysics of Maragha (RIAAM), Iran, under research project 
No.1/4165-14. F. Rahaman and A. Pradhan acknowledge 
IUCAA, 
Pune, India for awarding Visiting Associateship. The work was partly supported by VEGA 
Grant No. 2/0009/16. 
S. Capozziello acknowledges financial support of INFN (iniziative specifiche TEONGRAV and 
QGSKY). 
R. Pincak would 
like to thank the TH division in CERN for hospitality. This work was partially  supported 
by the JSPS KAKENHI Grant Number JP 25800136 and the research-funds 
given by  Fukushima University (K. Bamba).
This article is also based upon work from COST action CA15117 (CANTATA), supported by 
COST (
European Cooperation in Science and Technology). 
\end{acknowledgments}

\begin{appendix}

\section{The   action components   in 
terms of couplings of parallel and anti-parallel spins} 
\label{Apptermsnew}

In this Appendix we use the  definitions (\ref{w4new}) in order to calculate the 
different terms of (\ref{ss5}) in terms of couplings of parallel and anti-parallel spins 
\cite{Sepehri:2016sjq,Sepehri:2016svv}. In particular, we obtain: 
\begin{eqnarray}
 &&\!\!\!\!\!\!\!\!\!\!\!\!\!\!\!\!\!\!\!\!\!\!\!\!\!\!\!\!\!\!\!
 \langle
 F^{abc},F_{abc}\rangle_{Free-Free} = 
A^{ab}i\sigma_{ij}^{2}\partial_{a}^{i}\psi^{j}_{b}+\sigma^{0}_{ij}\psi^{
\dag a,i}\psi^{j}_{a}-\sigma^{1}_{ij}\psi^{\dag a, i}\psi^{j}_{a}
\nonumber \\
 &&\ \ \ \ \ \ \ \ \ \ \ \ \ \ \ 
 +
 \sigma^{0}_{i'i}(\psi^{\dag a, 
i'}i\sigma^{0}_{i'j}\sigma^{1}_{jk}\partial^{a,j}\psi^{k}_{a})(\psi^{\dag  
i}_{a}i\sigma^{0}_{ij}\sigma^{1}_{jk}\partial^{a,j}\psi^{k}_{a})
\nonumber \\
 &&\ \ \ \ \ \ \ \ \ \ \ \ \ \ \ 
 +
 \sigma^{1}_{i'i}(\psi^{\dag a, 
i'}i\sigma^{0}_{i'j}\sigma^{1}_{jk}\partial^{a,j}\psi^{k}_{a})(\psi^{\dag  
i}_{a}i\sigma^{0}_{ij}\sigma^{1}_{jk}\partial^{a,j}\psi^{k}_{a})
\nonumber \\
 &&\ \ \ \ \ \ \ \ \ \ \ \ \ \ \ 
 -
 \sigma^{0}_{i'i}(\psi^{\dag a, 
i'}i\sigma^{1}_{i'j}\sigma^{1}_{jk}\partial^{a,j}\psi^{k}_{a})(\psi^{\dag  
i}_{a}i\sigma^{1}_{ij}\sigma^{1}_{jk}\partial^{a,j}\psi^{k}_{a}),
\label{apprel1}
 \end{eqnarray}
  and
  \begin{eqnarray}
 &&\!\!\!\!\!\!\!\!\!\!\!\!\!
 \langle 
\partial^{b}\partial^{a}X^{i},\partial_{b}\partial_{a}X^{i}\rangle=\varepsilon^{abc}
\varepsilon^{ade}(\partial_{b}\partial_{c}X^{i}_{\alpha})(\partial_{e}\partial_{d}X^{i}_{ 
\beta})=\nonumber \\
 &&\ \ \ \ \ \ \ \ \ \ \ \ \ \ \ \ \ \ \ \     
 \Psi^{\dag a,U}\langle
     F_{abc},F^{a'bc}\rangle\Psi_{a'}^{L}+\Psi^{\dag a,L}\langle
     F_{abc},F^{a'bc}\rangle\Psi_{a'}^{U}- \Psi^{\dag a,U}\langle
     F_{abc},F^{a'bc}\rangle\Psi_{a'}^{U}
     \nonumber \\
 &&\ \ \ \ \ \ \ \ \ \ \ \ \ \ \ \ \ \ \ \ 
 -\Psi^{\dag a,L}\langle
     F_{abc},F^{a'bc}\rangle\Psi_{a'}^{L}
     +
     \Psi^{\dag a,L}\Psi^{\dag d,U}\partial_{d}\partial^{d'}\langle
     F_{abc},F^{a'bc}\rangle\Psi_{a'}^{L}\Psi^{U}_{d'}\nonumber \\
 &&\ \ \ \ \ \ \ \ \ \ \ \ \ \ \ \ \ \ \ \ 
     -
     \Psi^{\dag a,L}\Psi^{\dag d,U}\partial_{d}\langle
     F_{abc},F^{a'bc}\rangle\Psi_{a'}^{L}
     -
     \Psi^{\dag a,U}\Psi^{\dag d,L}\partial_{d}\langle
     F_{abc},F^{a'bc}\rangle\Psi_{a'}^{U} 
 \nonumber \\
            &&\ \ \ \ \ \ \ \ \ \ \ \ \ \ \ \ \ \ \ \ 
           + \psi^{\dag i,U}\langle
               F_{ijk},F^{i'jk}\rangle\psi_{i'}^{L}+\psi^{\dag i,L}\langle
               F_{ijk},F^{i'jk}\rangle\psi_{i'}^{U}-
               \psi^{\dag i,U}\langle
               F_{ijk},F^{i'jk}\rangle\psi_{i'}^{U}
               \nonumber \\
 &&\ \ \ \ \ \ \ \ \ \ \ \ \ \ \ \ \ \ \ \ 
 -\psi^{\dag i,L}\langle
               F_{ijk},F^{i'jk}\rangle\psi_{i'}^{L}+
               \psi^{\dag i,L}\psi^{\dag m,U}\partial_{m}\partial^{m'}\langle
               F_{ijk},F^{i'jk}\rangle\psi_{i'}^{L}\psi^{U}_{m'}
               \nonumber \\
 &&\ \ \ \ \ \ \ \ \ \ \ \ \ \ \ \ \ \ \ \ 
             -  \psi^{\dag i,L}\psi^{\dag m,U}\partial_{m}\langle
               F_{ijk},F^{i'jk}\rangle\psi_{i'}^{L}- \psi^{\dag i,U}\psi^{\dag 
m,L}\partial_{m}\langle
               F_{ijk},F^{i'jk}\rangle\psi_{i'}^{U} 
               \nonumber \\
 &&\ \ \ \ \ \ \ \ \ \ \ \ \ \ \ \ \ \ \ \  +    \Psi^{\dag a,U}\langle
                         F_{abc},F^{i'bc}\rangle\psi_{i'}^{L}+\Psi^{\dag a,L}\langle
                         F_{abc},F^{i'bc}\rangle\psi_{i'}^{U}-
                         \Psi^{\dag a,U}\langle
                         F_{abc},F^{i'bc}\rangle\psi_{i'}^{U}
                         \nonumber \\
 &&\ \ \ \ \ \ \ \ \ \ \ \ \ \ \ \ \ \ \ \  -\Psi^{\dag a,L}\langle
                         F_{abc},F^{i'bc}\rangle\psi_{i'}^{L}+
                         \Psi^{\dag a,L}\Psi^{\dag d,U}\partial_{d}\partial^{i'}\langle
                         F_{abc},F^{j'bc}\rangle\psi_{j'}^{L}\psi^{U}_{i'} 
                         \nonumber \\
 &&\ \ \ \ \ \ \ \ \ \ \ \ \ \ \ \ \ \ \ \ 
                         -\Psi^{\dag a,L}\Psi^{\dag d,U}\partial_{d}\langle
                         F_{abc},F^{i'bc}\rangle\psi_{i'}^{L}- 
                         \Psi^{\dag a,U}\Psi^{\dag 
d,L}\partial_{d}\langle
                         F_{abc},F^{i'bc}\rangle\psi_{i'}^{U}.
                         \label{apprel2}
 \end{eqnarray}
 Additionally, since $(\psi^{U}_{a})^{2}=0$ and $(\psi^{L}_{a})^{2}=0$, we get
  \begin{eqnarray}
&&
F(X)=\Sigma_{j}X_{j}^{2}=\Sigma_{j}[\psi^{
U}_{a}A^{ab,j}\psi^{L}_{b}-\psi^{L}_{a}A_{ab,j}\psi^{U}_{b}]^{2}
\nonumber \\ 
 &&\ \ \ \ \ \ \ \ \, 
=\Sigma_{j}(\psi^{U}_{a})^{2}(A^{ab,j})^{2}(\psi^{L}_{b})^{2}+(\psi^{L}_{a})^{2}(A_{ab,j}
)^{2}(\psi^{U}_{b})^{2}
\nonumber \\
 &&\ \ \ \ \ \ \ \ \ \ \  \
 -(\psi^{U}_{a}A^{ab,j}\psi^{L}_{b})(\psi^{L}_{a}A_{ab,j}\psi^{U}_{b
} )
 -(\psi^{L}_{a}A_{ab,j}\psi^{U}_{b})(\psi^{U}_{a}A^{ab,j}\psi^{L}_{b})=0
 \label{apprel3}
  \end{eqnarray}
  and
    \begin{equation}
\langle[X^{k},X^{i},X^{j}],[X_{k},X_{i},X_
{j}]\rangle=\Sigma_{n}\Sigma_{m}\alpha_{n+m}(\psi^{L})^{2n}(\psi^{U})^{2m}=0.
\label{apprel4}
 \end{equation}
 In the above expressions $a,b,c$ are indices of bound fermions, while $i,j,k$ are 
indices of free fermions, and $U,L$ refers to upper and lower spins. Additionally,    
 we have used the Pauli matrices definition as: $\sigma_{ij}^{1}=\Big{(}\begin{array}{cc}
0 & 1 \\
1 & 0
\end{array}\Big{)} $, $\sigma_{ij}^{0}=\Big{(}\begin{array}{cc}
1 & 0 \\
0 & 1
\end{array}\Big{)} $, $\sigma_{ij}^{2}=\Big{(}\begin{array}{cc}
 & -i \\
i & 0
\end{array}\Big{)} $.

 \section{ The expression of $\langle 
\partial^{b}\partial^{a}X^{i},\partial_{b}\partial_{a}X^{i}\rangle$ in terms of 
curvatures }
 \label{Bigfermcurvtermnew}

The term $\langle 
\partial^{b}\partial^{a}X^{i},\partial_{b}\partial_{a}X^{i}\rangle$ is expressed in terms 
of curvatures as:
 \begin{eqnarray}
 &&\!\!\!\!\!\!\!\!\!\!\!\!\!\!\!\!\!\!\!\!\!\!\!\! 
 \langle \partial^{b}\partial^{a}X^{i},\partial_{b}\partial_{a}X^{i}\rangle = 
 \Psi^{\dag a,U}
R_{aa'}^{anti-parallel} \Psi^{a',L}+\Psi^{\dag 
a,L}R_{aa'}^{anti-parallel}\Psi^{a',U}
\nonumber \\
&& \ \ \ \ \ \ \ \ \ \ \ \ \ \ \,
     - \Psi^{\dag a,U}R_{aa'}^{parallel}\Psi^{a',U}-\Psi^{\dag a,L}R_{aa'}^{parallel} 
\Psi^{a',L}
\nonumber \\
&& \ \ \ \ \ \ \ \ \ \ \ \ \ \ \, +\Psi^{\dag a,L}\Psi^{\dag 
d,U}\partial_{d}\partial^{d'}(R_{aa'}^{parallel}+R_{aa'}^{anti-
parallel})\Psi^{a',L}\Psi^{U}_{d'}
\nonumber \\
&& \ \ \ \ \ \ \ \ \ \ \ \ \ \ \, 
     - \Psi^{\dag a,L}\Psi^{\dag 
d,U}\partial_{d}(R_{aa'}^{parallel}+R_{aa'}^{anti-parallel}) \Psi^{a',L}
\nonumber \\
&& \ \ \ \ \ \ \ \ \ \ \ \ \ \ \, -\Psi^{\dag a,U}\Psi^{\dag 
d,L}\partial_{d}(R_{aa'}^{parallel}+R_{aa'}^{anti-parallel})\Psi^{
a',U} 
\nonumber \\
&& \ \ \ \ \ \ \ \ \ \ \ \ \ \ \,
+  \psi^{\dag i,U}R_{ii'}^{anti-parallel} 
\psi^{i',L}+\psi^{\dag 
i,L}R_{ii'}^{
anti-parallel}\psi^{i',U}
\nonumber \\
&& \ \ \ \ \ \ \ \ \ \ \ \ \ \ \, -\psi^{\dag 
i,U}R_{ii'}^{parallel}\psi^{i',U}-\psi^{\dag 
i,L}R_{ii'}^{
parallel}\psi^{i',L}
\nonumber \\
&& \ \ \ \ \ \ \ \ \ \ \ \ \ \ \,
                +
                \psi^{\dag i,L}\psi^{\dag 
m,U}\partial_{m}\partial^{m'}(R_{ij'}^{parallel}+R_{ij'}
^{anti-parallel})\psi^{i',L}\psi^{U}_{m'}
\nonumber \\
&& \ \ \ \ \ \ \ \ \ \ \ \ \ \ \, -
                \psi^{\dag i,L}\psi^{\dag 
m,U}\partial_{m}(R_{ij'}^{parallel}+R_{ij'}^{anti-
parallel})\psi^{i',L}
\nonumber \\
&& \ \ \ \ \ \ \ \ \ \ \ \ \ \ \,
                -
                \psi^{\dag i,U}\psi^{\dag 
m,L}\partial_{m}(R_{ij'}^{parallel}+R_{ij'}^{anti-
parallel})\psi^{i',U} 
\nonumber \\
&& \ \ \ \ \ \ \ \ \ \ \ \ \ \ \,
+
                      \Psi^{\dag 
a,U}R_{ai'}^{anti-parallel}\psi^{i',L}+\Psi^{\dag 
a,L}R_{
ai'}^{anti-parallel} \psi^{i',U}
\nonumber \\
&& \ \ \ \ \ \ \ \ \ \ \ \ \ \ \, -
                          \Psi^{\dag a,U}R_{ai'}^{parallel}\psi^{i',U}-\Psi^{\dag 
a,L}R_{ai'}^{
parallel}\psi^{i',L}
\nonumber \\
&& \ \ \ \ \ \ \ \ \ \ \ \ \ \ \, +
                          \Psi^{\dag a,L}\Psi^{\dag 
d,U}\partial_{d}\partial^{i'}(R_{ai'}^{
parallel}+R_{ai'}^{anti-parallel})\psi_{j'}^{L}\psi^{i',U}
\nonumber \\
  && \ \ \ \ \ \ \ \ \ \ \ \ \ \ \, -
         \Psi^{\dag a,L}\Psi^{\dag 
d,U}\partial_{d}(R_{ai'}^{parallel}+R_{ai'}^{
anti-parallel})\psi^{i',L}
\nonumber \\
&& \ \ \ \ \ \ \ \ \ \ \ \ \ \ \, -
     \Psi^{\dag a,U}\Psi^{\dag 
d,L}\partial_{d}(R_{ai'}^{parallel}+R_{ai'}^{
anti-parallel})\psi^{i',U}
\nonumber \\
&& \ \ \ \ \ \ \ \ \ \ \ 
  \approx
  ( R_{Free-Free}^{parallel})^{2}+( 
R_{Free-
Free}^{anti-parallel})^{2}+( R_{Free-Bound}^{parallel})^{2}
\nonumber \\
&& \ \ \ \ \ \ \ \ \ \ \ \ \ \ \,  +
(R_{Free-Bound}^{anti-
parallel})^{2}+( R_{Bound-Bound}^{parallel})^{2}+( 
R_{Bound-Bound}^{anti-parallel})^{2}
\nonumber \\
&& \ \ \ \ \ \ \ \ \ \ \ \ \ \ \, +
(R_{Free-Free}^{parallel} 
R_{Free-Free}^{anti-parallel})\,\partial^{2}(R_{Free-Free}^{parallel}+ R_
{Free-Free}^{anti-parallel} )
\nonumber \\
&& \ \ \ \ \ \ \ \ \ \ \ \ \ \ \,  +
(R_{Free-Bound}^{parallel} 
R_{Free-Bound}^{anti-
parallel})\,\partial^{2}(R_{Free-Bound}^{parallel}+ R_{Free-Bound}^{anti-parallel} 
)
\nonumber \\
&& \ \ \ \ \ \ \ \ \ \ \ \ \ \ \, 
+
(R_
{Bound-Bound}^{parallel} 
R_{Bound-Bound}^{anti-parallel})\,\partial^{2}(R_{Bound-Bound}^{parallel}+ R_
{Bound-Bound}^{anti-parallel} )
\label{w8},
 \end{eqnarray}
 where   $R_{Bound-Bound}^{anti-parallel}$ is the curvature produced by the interaction 
of 
two 
bound anti-parallel fermions, $R_{Bound-Bound}^{parallel}$ is the curvature created by 
interaction 
of two bound parallel fermions, $R_{Free-Free}^{anti-parallel}$ is the curvature 
produced 
by the 
interaction of two free anti-parallel fermions, $R_{Free-Free}^{parallel}$ is the 
curvature 
created by interaction of two free parallel fermions, $R_{Free-Bound}^{anti-parallel}$ 
is 
the 
curvature produced by the interaction of  free and bound anti-parallel fermions and 
$R_{Free-Free}
^{parallel}$ is the curvature created by interaction of  free and bound parallel 
fermions.

\section{The total action of $Mp$-branes in terms of the actions of $M1$-branes}
\label{actioncalculation}

In this Appendix we calculate the  total action of $Mp$-branes in terms of the actions of 
$M1$-branes. We begin with the Lagrangian of an $M1$-brane  in expression (\ref{ss5}), 
namely
\cite{Sepehri:2016jfx,Sepehri:2015eea,Sepehri:2016sjq,Sepehri:2016svv,Sepehri:2015ara}:
 \begin{eqnarray}
 &&
 L=\delta^{a_{1},a_{2}...a_{n}}_{b_{1}b_{2}....b_{n}}L^{b_{1}}_{a_{1}}...L^{b_{n}}_{a_{n}}
 \quad a,b,c=\mu,\nu,\lambda, 
 \end{eqnarray}
 where 
  \begin{eqnarray}
 && L^{b}_{a}=\delta_{a}^{b}\det (P_{abc}[ E_{mnl} +E_{mij}(Q^{-1}-\delta)^{ijk}E_{kln}]+
 \lambda F_{abc}),
 \label{Ldet}
\end{eqnarray}
with 
$F_{abc}=\delta_{a}^{b}\lambda\det(F)$. Substituting relations (\ref{w5new}), 
(\ref{w4new}), (\ref{w7new}),   (\ref{w8}) into the determinant of expression 
(\ref{Ldet}), we obtain:
 \begin{eqnarray}
&& \det(F)=\delta_{\rho\sigma}^{\mu\nu}\langle
 F^{\rho\sigma}\smallskip_{\lambda},F^{\lambda}\smallskip_{\mu\nu}\rangle
=\delta_{\rho\sigma}^{\mu\nu}\left(R^{anti-parallel, \rho\sigma}_{Free-Free,\mu\nu}+R^{
anti-
parallel, \rho\sigma}_{Bound-Bound,\mu\nu}+ 
R^{anti-parallel, \rho\sigma}_{Free-Bound,\mu\nu}\right)
\nonumber\\
&&
 \ \ \ \ \ \ \ \ \ \ \ \ \ \ \ \ \ \ \ \ \ \ \ \ \ \  \ \ \ \
 -
\delta_{\rho\sigma}^{
\mu\nu}\left(R^{parallel, \rho\sigma}_{Free-Free,\mu\nu}+
 R^{parallel, \rho\sigma}_{Bound-Bound,\mu\nu}+R^{parallel, 
\rho\sigma}_{Free-Bound,\mu\nu}
\right)\label{s12},
\end{eqnarray}
 \begin{eqnarray}
 &&
 \!\!\!\!\!\!\!\!\!\!\!\!\!\!
 \det(P_{abc}[ E_{mnl}
 +E_{mij}(Q^{-1}-\delta)^{ijk}E_{kln}])
  \nonumber\\
 &&\!\!\!\!\!\!\!\!
= \delta_{\rho\sigma}^{\mu\nu}
 \Biggl[\left(g^{\mu}_{\rho}g^{\nu}_{\sigma}+ \langle 
\partial^{b}\partial^{a}X^{i},\partial_{b}\partial_
{a}X^{i}\rangle+\cdots\right)
 -
 \frac{(g^{\mu}_{\rho}g^{\nu}_{\sigma}+
 \langle 
\partial^{b}\partial^{a}X^{i},\partial_{b}\partial_{a}X^{i}\rangle+...)}{[(\lambda)^{2} 
\det([X^{j}_{\alpha}T^{\alpha},X^{k}_{\beta}T^{\beta},X^{k'}_{\gamma}T^{\gamma}])]}\Biggr]
 \nonumber\\
 &&\!\!\!\!\!\!\!\!
 =
 \Biggl[( R_{Free-Free}^{parallel})^{2}+( 
R_{Free-Free}^{anti-parallel})^{2}+( R_{Free-
Bound}^{parallel})^{2}+\nonumber \\&&
 ( R_{Free-Bound}^{anti-parallel})^{2}+( R_{Bound-Bound}^{parallel})^{2}+( 
R_{Bound-Bound}^{anti-
parallel})^{2}+\nonumber \\&& (R_{Free-Free}^{parallel} 
R_{Free-Free}^{anti-parallel})\partial^{2}(
R_{Free-Free}^{parallel}+ R_{Free-Free}^{anti-parallel} )+\nonumber 
\\&&(R_{Free-Bound}^{parallel} 
R_{Free-Bound}^{anti-parallel})\partial^{2}(R_{Free-Bound}^{parallel}+ 
R_{Free-Bound}^{anti-
parallel} )+\nonumber \\&&(R_{Bound-Bound}^{parallel} 
R_{Bound-Bound}^{anti-parallel})\partial^{2}(
R_{Bound-Bound}^{parallel}+ R_{Bound-Bound}^{anti-parallel} 
)\Biggr]\left(1-\frac{1}{m_{g}^{2}}\right),
\label{s13}
 \end{eqnarray}
where 
$m_{g}^{2}=[(\lambda)^{2}\det([X^{j}_{\alpha}T^{\alpha},X^{k}_{\beta}T^{\beta},X^{k'} _{
\gamma}T^{\gamma}])]
 $ is the square of the graviton mass. Applying this relation, we can additionally obtain 
 \begin{eqnarray}
  &&\det(Q) \sim
-[(\lambda)^{2}\det([X^{j}_{\alpha}T^{\alpha},X^{k}_{\beta}T^{\beta},X^{k'}_{\gamma}T^{
\gamma}])]\det(g)=-m_{g}^{2}\det(g).
 \label{s14}
 \end{eqnarray}
Substituting the above relations into the action (\ref{ss5}), we obtain:
 \begin{eqnarray}
 && \!\!\!\!\!\!\!
 \!\!\!\!\!\!\!\!
 L_{b}^{a}=\delta_{b}^{a}
 \Biggl\{-(1-m_{g}^{2})\Bigl[( 
R_{Free-Free}^{parallel})^{2}+( R_{
Free-Free}^{anti-parallel})^{2}+
 ( R_{Free-Bound}^{parallel})^{2}
 \nonumber \\
 && \ \ \ \ \ \ \ \ \ \ \ \ \ \ \ \ \ \ \ \
 +
 ( R_{Free-Bound}^{anti-parallel})^{2}+( R_{Bound-Bound}^{parallel})^{2}+( 
R_{Bound-Bound}^{anti-
parallel})^{2}
\nonumber \\
 && \ \ \ \ \ \ \ \ \ \ \ \ \ \ \ \ \ \ \ \
+
 (R_{Free-Free}^{parallel} 
R_{Free-Free}^{anti-parallel})\partial^{2}(R_{Free-Free}^{parallel}+ R_{
Free-Free}^{anti-parallel} )
\nonumber\\
 && \ \ \ \ \ \ \ \ \ \ \ \ \ \ \ \ \ \ \ \
+
(R_{Free-Bound}^{parallel} 
R_{Free-Bound}^{anti-parallel})\partial^{2}(R_{Free-Bound}^{parallel}+ 
R_{Free-Bound}^{anti-parallel})
\nonumber\\
 && \ \ \ \ \ \ \ \ \ \ \ \ \ \ \ \ \ \ \ \
+
(R_{Bound-Bound}^{parallel} 
R_{Bound-Bound}^{anti-parallel})\partial^{2}(R_{Bound-Bound}^{parallel}
+ R_{Bound-Bound}^{anti-parallel} )\Bigr]\nonumber\\
&& \ \ \ \ \,
+ 
m_{g}^{2}\lambda^{2}\Big[\delta_{\rho\sigma}^{\mu\nu}(R^{anti-parallel, \rho\sigma}_{
Free-Free,\mu\nu}+R^{anti-parallel, \rho\sigma}_{Bound-Bound,\mu\nu}+
 R^{anti-parallel, \rho\sigma}_{Free-Bound,\mu\nu})
 \nonumber\\
 &&\ \ \ \ \ \ \ \ \ \ \ \ \ \ \ \,
 -
 \delta_{\rho\sigma}^{\mu\nu}(R^{parallel, \rho\sigma}_{Free-Free,\mu\nu}+ 
R^{parallel, \rho\sigma}_{Bound-Bound,\mu\nu}+R^{parallel, \rho\sigma}_{Free-Bound,\mu\nu}
)\Big] 
\Biggr\}.
 \label{s15}
 \end{eqnarray} 
 
As a next step, in order to calculate the total action of an $Mp$-brane, we need to 
obtain 
the sum over curvatures produced by free and bound fermions which are attached to the 
$M1$-branes that construct the $Mp$-brane. In particular, $S_{Mp} = - \int d^{p+1}x 
\L_{Mp} $, where 
 \begin{eqnarray}
  && 
\L_{Mp}=\det(M)=\sum_{n=1}^{p}\delta^{a_{1},a_{2}...a_{n}}_{b_{1}b_{2}....b_{n}}L^{b_{1}}_
{a_{1}}..
.L^{b_{n}}_{a_{n}},
 \end{eqnarray}
 with $\delta^{a_{1},a_{2}...a_{n}}_{b_{1}b_{2}....b_{n}} 
\delta^{\rho_{1}\sigma_{1}}_{\mu_{1}\nu_{1}}...\delta^{\rho_{p}\sigma_{p}}_{\mu_{p}\nu_{p}
} 
=\delta^{\rho_{1}\sigma_{1}...\rho_{p}\sigma_{p}}_{\mu_{1}\nu_{1}...\mu_{p}\nu_{p}}$. 
Moreover, we have $ L_{M1,i}=L^{b_{i}}_{a_{i}}=\det(M_{i})\sim M_{i}$ and 
$\sqrt{-g}= \sqrt{-\det(g_{1})\det(g_{2}
)...\det(g_{
p})}$. Hence, assembling everything, we find 
\begin{eqnarray}
 &&
 \!\!\!\!\!\!
 S_{Mp} = 
  - \int d^{p+1}x\sqrt{-g}
  \Biggl\{
  \Biggl\{
  \sum_{n=1}^{p}(1-m_{g}^{2})^{n}
 \Big[
 ( R_{Free-Free}^{parallel})^{2}+( R_{Free-Free}^{anti-parallel})^{2}+( 
R_{Free-Bound}^{parallel})
^{2}
\nonumber 
\\
&& \ \ \ \ \ \ \ \ \ \ \ \ \ \ \ \ \ \ \ \ \ \ \ \ 
+ ( R_{Free-Bound}^{anti-parallel})^{2}+( R_{Bound-Bound}^{parallel})^{2}+( 
R_{Bound-Bound}^{anti-
parallel})^{2}
\nonumber
\\
&& \ \ \ \ \ \ \ \ \ \ \ \ \ \ \ \ \ \ \ \ \ \ \ \ 
+
 (R_{Free-Free}^{parallel} 
R_{Free-Free}^{anti-parallel})\partial^{2}(R_{Free-Free}^{parallel}+ R_{
Free-Free}^{anti-parallel} )
\nonumber
\\
&& \ \ \ \ \ \ \ \ \ \ \ \ \ \ \ \ \ \ \ \ \ \ \ \ 
+(R_{Free-Bound}^{parallel} 
R_{Free-Bound}^{anti-parallel}
)\partial^{2}(R_{Free-Bound}^{parallel}+ R_{Free-Bound}^{anti-parallel} )
\nonumber 
\\
&& \ \ \ \ \ \ \ \ \ \ \ \ \ \ \ \ \ \ \ \ \ \ \ \ 
+(R_{Bound-
Bound}^{parallel} 
R_{Bound-Bound}^{anti-parallel})\partial^{2}(R_{Bound-Bound}^{parallel}+ 
 R_{Bound-Bound}^{anti-parallel} )
 \Big]^{n}
 \nonumber
 \\
 &&\ \ \ \ \ \,
 + 
\sum_{n=1}^{p}m_{g}^{2n}\lambda^{2n}\delta_{\rho_{1}\sigma_{1}\cdots\rho_{n}\sigma_{n}}^{
\mu_
{1}\nu_{
1}\cdots
\mu_{n}\nu_{n}} 
\Big(
R^{anti-parallel, \rho_{1}\sigma_{1}}_{Free-Free,\mu_{1}\nu_{1}}+R^{anti-parallel,
\rho_ { 1
}\sigma_{
1}}_{Bound-Bound,\mu_{1}\nu_{1}}+ 
R^{anti-parallel, \rho_{1}\sigma_{1}}_{Free-Bound,\mu_{1}\nu_{1}}
\nonumber\\
&& \ \ \ \ \ \ \ \ \ \ \ \ \ \ \ \ \ \ \ \ \ \ \ \ \ \ \ \ \ \ \ \ \ \ \ \ \ \ \
-R^{
parallel, \rho_{
1}\sigma_{1}}_{Free-Free,\mu_{1}\nu_{1}}-
R^{parallel, \rho_{1}\sigma_{1}}_{Bound-Bound,\mu_{1}\nu_{1}}-R^{parallel, 
\rho_{1}\sigma_{1
}}_{Free-
Bound,\mu_{1}\nu_{1}}\Big)
\nonumber\\
&& \ \ \ \ \ \ \ \ \ \ \    \ \ \  \times\cdots
\cdots
\times
\Bigl(
R^{anti-parallel, \rho\sigma}_{Free-Free,\mu\nu}+R^{anti-parallel, \rho\sigma}_{
Bound-Bound,\mu\nu}+
 R^{anti-parallel, \rho_{n}\sigma_{n}}_{Free-Bound,\mu_{n}\nu_{n}}
 \nonumber\\
 &&  \ \ \ \ \ \ \ \ \ \ \  \ \ \ \ \ \ \ \ \ \  \ \ 
-R^{parallel, \rho_{n}\sigma_{n}}_{Free-Free,\mu_{n}\nu_{n}}-R^{parallel, 
\rho_{n}\sigma_{n}
 }_{Bound-
Bound,\mu_{n}\nu_{n}}-
R^{parallel, \rho_{n}\sigma_{n}}_{Free-Bound,\mu_{n}\nu_{n}}\Bigr)
\Biggr\}^{\frac{1}{2}}
\Biggr\}.
\label{s17app}
 \end{eqnarray}
Hence, we  have calculated the  total action of $Mp$-branes in terms of the actions of 
$M1$-branes, i.e in terms of the individual curvatures.

\section{The total action of compactified $Mp$-branes in terms of the actions of 
$M1$-branes} 
\label{compactified}

In this Appendix we repeat the calculation of the previous Appendix, but in the case of 
compactified branes of Section \ref{Sectioncontraction}.  We make the choice 
${\displaystyle 
<\psi^{3}>=\frac{r}{l_{s}^{3/2}}}
$ where $l_{s}$ is
the string coupling, $r$ is radius of a circle that branes are compactified on it. Using 
this definition relations 
(\ref{apprel1}),(\ref{apprel2}),(\ref{apprel3}),(\ref{apprel4}), using also 
$[X^{a},X^{b},X^{c}]=F^{abc}$, $ X^{b}X^{c}=\psi_{L}^{b}\psi_{U}^{c}$ and 
$[X^{a},\psi_{L}^{b}]=\partial_{a,U}\psi_{L}^{b}$ give:
\begin{eqnarray}
&& 
\Sigma_{a,b,c=0}^{3} \langle
F^{abc},F_{abc}\rangle 
 = 
6\left(\frac{r^{2}}{l_{s}^{3}}\right)\Sigma_{a,b=0}^{2}\partial_{U}^{a}\Psi_{L,b}\partial_
{a,U}\Psi_{L}^{b}
+6\left(\frac{r^{2}}{l_{s}^{3}}\right)\Sigma_{a,b=0}^{2}\partial_{L}^{a}\Psi_{U,b}
\partial_{a,L}\Psi_{U}^{b}\nonumber \\
&&
-12\left(\frac{r^{2}}{l_{s}^{3}}\right)\Sigma_{a,b=0}^{2}\partial_{U}^{a}\Psi_{L,b}
\partial_{a,L}\Psi_{U}^{b}
 +E_{Extra} ,
 \label{ss24}
 \end{eqnarray}
 and
 \begin{eqnarray}
\partial_{a}\partial_{b}X^{i}\partial^{a}\partial^{b}X^{i}
&=& 
6\left(\frac{r^{2}}{l_{s}^{3}}\right)\Sigma_{a,i=0}^{2}\partial_{U}^{a}\Psi_{L,i}\partial_
{a,U}\
Psi_{L}^{i}
+6\left(\frac{r^{2}}{l_{s}^{3}}\right)\Sigma_{a,i=0}^{2}\partial_{L}^{a}\Psi_{U,i}
\partial_{a,L}\
Psi_{U}^{i}\nonumber \\
&-&12\left(\frac{r^{2}}{l_{s}^{3}}\right)\Sigma_{a,i=0}^{2}\partial_{U}^{a}\Psi_{L,i}
\partial_{a,L}\
Psi_{U}^{i}
 +E_{Extra},
 \label{sss24}
\end{eqnarray}
where $ E_{Extra}=- 
6\Sigma_{a,b,c,a',b',c'=0,\neq3}^{2}\varepsilon_{abcD}\varepsilon_{a'b'c'}^{D}X^{a}X^{b}
\psi_{L}^{
b}\psi_{L}^{b'} $.
Expression (\ref{ss24}) indicates  that by compacting branes the sign of gravity changes 
and repulsive gravity emerges. In fact, the symmetry of the system is broken again and 
terms related to attractive force are canceled. This reveals that the repulsive 
gravity between parallel spins is the underlying reason for the appearance of Pauli 
exclusion principle. 

Let us now calculate  the curvatures of the system. From (\ref{curvauxil}), 
(\ref{w7new}) we find:
\begin{eqnarray}
  && R_{MN}=\langle
 F^{\rho}\smallskip_{\sigma\lambda},F^{\lambda}\smallskip_{\mu\nu}\rangle= 
6\left(\frac{r^{2}}{l_{s}^{3}}\right)\Sigma_{\rho,\mu=0}^{2}\partial_{U}^{\rho}\psi_{L,
\sigma}\partial_{U,\mu}\psi_{L,\nu}+
 6\left(\frac{r^{2}}{l_{s}^{3}}\right)\Sigma_{\rho,\mu=0}^{2}\partial_{L}^{\rho}\Psi_{U,
\sigma}\partial_{L,\mu}\Psi_{U,\nu}
\nonumber\\&&
-
  12\left(\frac{r^{2}}{l_{s}^{3}}\right)\Sigma_{\rho,\mu=0}^{2}\partial_{L}^{\rho}\Psi_{U,
\sigma}\partial_{U,\mu}\Psi_{L,\nu}=
   R_{MN,parallel}-R_{MN,anti-parallel}.
\label{s25}
 \end{eqnarray}
 while
 \begin{eqnarray}
 &&
 \!\!\!\!\!\!\!\!\!\!\!\!\!
 \partial_{a}\partial_{b}X^{i}\partial^{a}\partial^{b}X^{i}=
R^{parallel}_{Free-Free}+R^{parallel}_{Free-Bound}+R^{parallel}_{Bound-Bound}
\nonumber\\
&& \ \ \ \ \ \ \ \ \ \ \ \ 
\ \ \ \ \ 
-R^{anti-parallel}_{Free-Free}-R^{anti-parallel}_{Free-Bound}-R^{anti-parallel}_{
Bound-Bound}.
\label{ss26}
  \end{eqnarray}

 Substituting relations (\ref{w5new}), (\ref{w4new}), (\ref{w7new}),   (\ref{ss26}) into 
the determinant of expression (\ref{Ldet}), we obtain:
\begin{eqnarray}
 &&\!\!\!\!\!\!\!\!\!\!\!\!\!\!\!\!\!
 \det(F)=\delta_{\rho\sigma}^{\mu\nu}\langle
 F^{\rho\sigma}\smallskip_{\lambda},F^{\lambda}\smallskip_{\mu\nu}\rangle
 =6\left(\frac{r^{2}}{l_{s}^{3}}\right)\delta_{\rho\sigma}^{\mu\nu} 
\Big(R^{parallel, \rho\sigma}_{Free-Free,\mu\nu}+R^{parallel, \rho\sigma}_{Bound-Bound,
\mu\nu}
 \nonumber\\
 &&
 \ \ \  \ +
 R^{parallel, \rho\sigma}_{Free-Bound,\mu\nu} -
 R^{anti-parallel, \rho\sigma}_{Free-Free,
\mu\nu}-R^{anti-parallel, \rho\sigma}_{Bound-Bound,\mu\nu}-
 R^{anti-parallel, \rho\sigma}_{Free-Bound,\mu\nu})
 \Big),
 \label{ss27bc}
 \end{eqnarray}
and
 \begin{eqnarray}
 &&
 \!\!\!\!\!\!\!\!\!\!\!\!\!\!
 \det(P_{abc}[ E_{mnl}
 +E_{mij}(Q^{-1}-\delta)^{ijk}E_{kln}])
  \nonumber\\
 &&\!\!\!\!\!\!\!\!
= \delta_{\rho\sigma}^{\mu\nu}
 \Biggl[\left(g^{\mu}_{\rho}g^{\nu}_{\sigma}+ \langle 
\partial^{b}\partial^{a}X^{i},\partial_{b}\partial_
{a}X^{i}\rangle+\cdots\right)
 -
 \frac{(g^{\mu}_{\rho}g^{\nu}_{\sigma}+
 \langle 
\partial^{b}\partial^{a}X^{i},\partial_{b}\partial_{a}X^{i}\rangle+...)}{[(\lambda)^{2} 
\det([X^{j}_{\alpha}T^{\alpha},X^{k}_{\beta}T^{\beta},X^{k'}_{\gamma}T^{\gamma}])]}\Biggr]
 \nonumber\\
 &&\!\!\!\!\!\!\!\!
 =
 \left(\frac{r^{2}}{l_{s}^{3}}\right) \left(1-\frac{1}{m_{g}^{2}}
\right)\Big(R^{parallel}_{Free-Free}+R^{parallel}_{Free-Bound}+R^{parallel}_{Bound-Bound}
\nonumber\\
&& \ \ \ \ \ \ \ \ \ \ \ \ \ \ \ \ \ \ \ \ \ \ \ \  \,
-R^{anti-parallel}_{Free-Free}-R^{anti-parallel}_{Free-Bound}-R^{anti-parallel}_{
Bound-Bound}\Big),
\label{s13bc}
 \end{eqnarray}
where 
$m_{g}^{2}=[(\lambda)^{2}\det([X^{j}_{\alpha}T^{\alpha},X^{k}_{\beta}T^{\beta},X^{k'} _{
\gamma}T^{\gamma}])]
 $ is the square of the graviton mass. Applying this relation, we can again obtain 
(\ref{s14}). Substituting the above relations into the action (\ref{ss5}), and setting  
for simplicity $\frac{6r^{2}}{l_{s}^{3}}=1$, we acquire:
 \begin{eqnarray}
 && \!\!\!\!\!\!\!
 \!\!\!\!\!\!\!\ 
L_{b}^{a}=\delta_{b}^{a}
\Bigl
(1-m_{g}^{2})\Big(R^{parallel}_{Free-Free}+R^{parallel}_{Free-Bound}+R^{parallel}_{
Bound-Bound}
\nonumber
\\
&& \ \ \ \ \ \ \ \ \ \ \ \ \ \ \ \ \ \ \ 
- 
R^{anti-parallel}_{Free-Free}-R^{anti-parallel}_{Free-Bound}-R^{anti-parallel}_{
Bound-Bound}\Big)
\nonumber\\
 &&  \ \
  + 
m_{g}^{2}\lambda^{2}\delta_{\rho\sigma}^{\mu\nu}
\Big(R^{parallel, \rho\sigma}_{Free-Free,
\mu\nu}+R^{parallel,
\rho\sigma}
_{Bound-Bound,\mu\nu}+ 
R^{parallel, \rho\sigma}_{Free-Bound,\mu\nu}
\nonumber
\\
&& \ \ \ \ \ \ \ \ \ \ \ \ \ \ \ \ \ \ 
-
R^{anti-parallel, \rho\sigma}_{Free-Free,
\mu\nu}-R^{anti-parallel, \rho\sigma}_{Bound-Bound,\mu\nu}-
 R^{anti-parallel, \rho\sigma}_{Free-Bound,\mu\nu}\Big).
 \label{ss29bc}
 \end{eqnarray}
This Lagrangian indicates that by compacting the branes the sign of curvatures reverses, 
and hence the gravity behavior changes from attractive to repulsive, and thus  the branes 
tend to mutual remove.

As a next step, in order to calculate the total action of an $Mp$-brane, we need to 
obtain the sum over curvatures produced by free and bound fermions which are attached to 
the $M1$-branes that construct the $Mp$-brane. In particular, $S_{Mp} = - \int d^{p+1}x 
\L_{Mp} $, where 
 \begin{eqnarray}
  && 
\L_{Mp}=\det(M)=\sum_{n=1}^{p}\delta^{a_{1},a_{2}...a_{n}}_{b_{1}b_{2}....b_{n}}L^{b_{1}}_
{a_{1}}..
.L^{b_{n}}_{a_{n}},
 \end{eqnarray}
 with $\delta^{a_{1},a_{2}...a_{n}}_{b_{1}b_{2}....b_{n}} 
\delta^{\rho_{1}\sigma_{1}}_{\mu_{1}\nu_{1}}...\delta^{\rho_{p}\sigma_{p}}_{\mu_{p}\nu_{p}
} 
=\delta^{\rho_{1}\sigma_{1}...\rho_{p}\sigma_{p}}_{\mu_{1}\nu_{1}...\mu_{p}\nu_{p}}$. 
Moreover, we have $ L_{M1,i}=L^{b_{i}}_{a_{i}}=\det(M_{i})\sim M_{i}$ and 
$\sqrt{-g}= \sqrt{-\det(g_{1})\det(g_{2}
)...\det(g_{
p})}$. Hence, assembling everything, we find 
 \begin{eqnarray}
 S_{Mp}  
 &=& - 
 \int 
d^{p+1}x\sqrt{-g}
\Biggl\{
\Biggl\{
\sum_{n=1}^{p}\Big(1-m_{g}^{2})
\Big(R^{parallel}_{Free-Free}+R^{parallel}_{
Free-Bound}
+R^{parallel}_{Bound-Bound}
\nonumber
\\
&&
 \ \ \ \ \ \ \ \ \ \ \  \ \ \ \ \ \ \ \ \ \  \ \  \ \ \ \  \ \ \ \ \  \ \ \ \ \ \ \ \ \ \ 
 \ \ 
-R^{anti-parallel}_{Free-Free}-R^{anti-parallel}_{Free-Bound}-R^{anti-parallel}_{
Bound-Bound}\Big)^{n}
\nonumber\\
&&
+
\sum_{n=1}^{p}m_{g}^{2n}\lambda^{2n}\delta^{\rho_{1}\sigma_{1}...\rho_{n}\sigma_{n}}_{\mu_
{1}\nu_{
1}...\mu_{n}\nu_{n}} 
\Big(R^{parallel, \rho_{1}\sigma_{1}}_{Free-Free,\mu_{1}\nu_{1}}+R^{parallel, \rho_{1}
\sigma_ { 1 }}_{ Bound-Bound,
\mu_{1}\nu_{1}}+
 R^{parallel, \rho_{1}\sigma_{1}}_{Free-Bound,\mu_{1}\nu_{1}}
 \nonumber\\ 
&&
 \ \ \ \ \ \ \ \ \ \ \  \  \ \ \ \  \ \ \ \ \ \ \ \ \ \ 
 \ \ 
 -R^{anti-parallel, \rho_{n}\sigma_{1}}_{Free-Free,\mu_{1}\nu_{1}}
-R^{anti-parallel, \rho_{1}\sigma_{1}}_{Bound-Bound,
\mu_{1}\nu_{1}}-
R^{anti-parallel, \rho_{1}\sigma_{1}}_{Free-Bound,\mu_{1}\nu_{1}}\Big)
\nonumber\\
&& \ \ \ \ \ \ \ \ \  \times\cdots
\cdots
\times
 \Big(R^{parallel, \rho_{1}\sigma_{n}}_{Free-Free,\mu_{n}\nu_{n}}+ 
R^{parallel, \rho_{n}\sigma_{n}}_{Bound-Bound,\mu_{n}\nu_{n}}+R^{parallel, 
\rho_{n}\sigma_{n } }_{ Free-Bound,
\mu_{n}\nu_{
n}}
\nonumber\\
&&\ \ \  \  \ \ \ \  \ \ \  \ 
-
R^{anti-parallel,\rho_{n}\sigma_{n}}_{Free-Free,\mu_{n}\nu_{n}}-R^{anti-parallel,\rho_{n}
\sigma_{n}}_{ Bound-Bound,
\mu_{n}\nu_{n}}- 
R^{anti-parallel, \rho_{n}\sigma_{n}}_{Free-Bound,\mu_{n}\nu_{n}}\Big)
\Biggr\}^{\frac{1}{2}}
\Biggr\}.
\label{ss30bc}
 \end{eqnarray}
Hence, we  have calculated the  total action of compactified $Mp$-branes in terms of 
the actions of $M1$-branes, i.e in terms of the individual curvatures.

\end{appendix}

\end{document}